\documentclass[
 preprint,
 amsmath,amssymb,
 aip, jcp, floatfix,
]{revtex4-2}

\usepackage[english]{babel}
\usepackage[margin=0.8in]{geometry}
\usepackage{graphicx}
\usepackage{bm}
\usepackage[colorlinks=true, allcolors=blue]{hyperref}
\usepackage{indentfirst}
\usepackage{wrapfig}
\DeclareUnicodeCharacter{2212}{\textendash}

\begin{document}

\title{Size-dependent second-order-like phase transitions in Fe nanocluster melting from low-temperature structural isomerization}

\author{Louis E. S. Hoffenberg}
    \affiliation{Department of Chemical and Biological Engineering, Princeton University, Princeton, New Jersey 08540}
\author{Alexander Khrabry}
    \altaffiliation{Andlinger Center for Energy and the Environment, Princeton University, Princeton, New Jersey 08540}
\author{Yuri Barsukov}
    \altaffiliation{Princeton Plasma Physics Laboratory, Princeton, New Jersey 08540}
\author{Igor D. Kaganovich}
    \altaffiliation{Princeton Plasma Physics Laboratory, Princeton, New Jersey 08540}
\author{David B. Graves}
    \email{dgraves@princeton.edu}

\date{\today}

\begin{abstract}
In this work, the melting phase transitions of $Fe_{n}$ nanoclusters with $10 \leq n \leq 100$ atoms are investigated using classical many-body molecular dynamics simulations. For many cluster sizes, surface melting occurs at much lower temperatures than core melting. Surface and core melting points, and energetic melting points (temperatures of maximum heat capacity, $C_v$) are calculated for all cluster sizes. Melting properties are found to be strong functions of cluster structure. Cluster sizes with closed-shell structures always have first-order-like phase transitions. Almost one-third of cluster sizes in the analyzed range exhibit second-order-like phase transitions due to the presence of multiple structural configurations close in energy. 1-shell clusters with one to a few more atoms than a neighboring closed-shell structure have very low surface melting points and very high energetic melting points compared to their closed-shell counterparts. In clusters above 50 atoms with certain core structures, melting of the surface before the core was observed.
\end{abstract}

\maketitle

\section{Introduction}
\label{intro}

Nanoparticles (NPs) – i.e., particulate material with characteristic dimensions under 100 nm – have interesting properties that make them desirable for catalytic\cite{tao_metal_2014,imaoka_magic_2013}, optoelectronic\cite{pescaglini_metal_2015}, and biomedical\cite{mcnamara_nanoparticles_2017} applications, among many others. These properties can depend on many factors like nanoparticle size, composition, degree of crystallinity, and other structural elements. The phase of the NP can have a significant effect on properties since atoms in liquid particles have considerably more mobility than in solid form. For example, it has been proposed that carbon nanotube (CNT) growth on catalytic iron NPs depends in part on carbon precursor adsorption, surface diffusion, and dissolution into the NP\cite{puretzky_situ_2005, helveg_atomic-scale_2004} – all of which can be influenced by the NP phase\cite{amara_modeling_2017}. The transition between solid and liquid NP phases is therefore of potential importance in multiple applications.

The melting points for NPs are known to differ from those of their corresponding bulk materials\cite{nanda_size-dependent_2009,Rare_gas_Beck,Solid–liquid_Jellinek}. NP melting temperatures are lower than the bulk melting point due to larger surface-atom-to-volume-atom ratios\cite{pawlow_ober_1909}. Surface atoms are bonded to fewer atoms than the inner atoms are, so smaller NPs require less energy to melt than larger ones. Their melting points scale according to the Gibbs-Thomson equation:
\begin{equation}
    T_{m, NP} = T_{m, bulk} \left(1 - 2 \frac{\sigma_{sl}}{\Delta H_{m} \rho_s r}\right)
\end{equation}
where $\sigma_{sl}$ is the solid-liquid interfacial energy, $\Delta H_{m}$ is the bulk latent heat of the melting, $\rho$ is the bulk solid density, and $r$ is the radius of the NP\cite{jackson_melting_1990}.

It is also known that when the NP radius is below some threshold size range (i.e., approaching the nanocluster regime), melting points no longer follow the Gibbs-Thomson equation. Instead, melting temperatures can fluctuate strongly with cluster size, with the addition or subtraction of a single atom. Sometimes, the nanocluster melting points can exceed those of the bulk solid\cite{breaux_hot_2003}.

This variation in melting points is attributed to interrelated geometric and electronic quantum size effects\cite{schmidt_irregular_1998,susan_correlation_2013,joshi_magic_2006}, collectively called magic number effects. Magic number \textit{clusters} are nanoclusters with particularly stable structures due to either configurational symmetry that maximizes bonding between atoms (corresponding to geometric magic numbers) or electronic effects that stabilize certain cluster geometries (i.e., electronic magic numbers)\cite{schmidt_irregular_1998,susan_correlation_2013,joshi_magic_2006}. Magic numbers have been documented to affect melting behavior and the Gibbs free energies of formation for small metal clusters\cite{girshick_homogeneous_2009,li_free_2007}.

In larger systems and NPs, atoms are frequently approximated as \textit{surface} or \textit{bulk}. The transition regime between Gibbs-Thomson NP scaling and the nanocluster fluctuation regime is thought to occur when the cluster's atoms can no longer fit the surface-bulk binary status\cite{gaston_cluster_2018}. When clusters fall in this regime, there are fewer bulk atoms in general, and there are distinct \textit{types} of surface atoms with different binding energies which determine cluster energetics. Because certain structures have perfectly closed atomic shells (geometric magic numbers)\cite{martin_shells_1996}, nanoclusters with more or fewer atoms have notably different binding energies per atom and therefore different melting temperatures\cite{bagrets_lowering_2010,susan_correlation_2013,joshi_magic_2006}. Ion calorimetry measurements of Al nanoclusters have revealed that the transition between Gibbs-Thomson NP scaling and nanocluster variation of melting temperatures occurs between clusters of 150 and 342 atoms\cite{yalamanchali_melting_2017}. Simulations of Ni nanoclusters observed the Gibbs-Thomson NP scaling in clusters as small as 90 atoms\cite{neyts_numerical_2009}.

The nanocluster size range, which is the focus of this work, is particularly interesting for some catalysis applications such as the catalytic growth of carbon nanotubes with Fe or Fe-containing alloys in floating catalyst chemical vapor deposition (FCCVD)\cite{amara_modeling_2017,sehrawat_floating_2024}. Fe nanoclusters of up to $\sim$100 atoms ($\sim$1.2 nm in diameter) are most relevant to the growth of single-wall CNTs (SWCNTs)\cite{amara_modeling_2017}. This work focuses on Fe nanoclusters. Despite interest in iron nanoparticles for the catalytic growth of CNTs, among other applications, the melting behavior of Fe nanoclusters has garnered few dedicated studies\cite{kim_magic_2014,diep_melting_1989,ding_size_2004}. Furthermore, no study has analyzed the majority of the Fe nanocluster size range, let alone the relationship between cluster structure and mechanisms of the melting phase transition, where we distinguish between the melting of the surface, internal layers, and the core. Caloric curves describing the energetics of Fe cluster melting behavior in this work are used in the accompanying paper \textit{Khrabry et al.}\cite{khrabry2024gibbsfreeenergiesfe} to determine the free energies of cluster formation for kinetic modeling of nucleation and growth from condensing vapor.

The process of melting in nanoclusters differs from bulk material melting and NP melting. Bulk melting is described by a sharp increase in atomic mobility of all atoms\cite{lindemann1910} and a steep rise in a caloric curve (a graph of cluster energy vs. temperature) at the melting point, indicating a first-order phase transition. NP melting, on the other hand, is often characterized by surface melting followed by melting of the NP core\cite{nanda_size-dependent_2009, huang_two-stage_2012}. For small nanoclusters, the process of ''melting`` involves a dynamic coexistence between ordered and disordered phases, a phenomenon generally not seen in nanoparticles or bulk materials\cite{rice_evolution_2009,schebarchov_solid-liquid_2006,aguado_neither_2016}.

Because NP and nanocluster melting generally occurs on length scales and time scales that are difficult to resolve experimentally, molecular simulation is often employed. Monte Carlo (MC) methods can efficiently sample configurational potential energy surfaces and construct caloric curves to describe cluster phase transitions\cite{srinivas_ab_2000,frantz_magic_2001,yurtsever_many-body_2000}. Molecular dynamics (MD) simulations construct time trajectories of atoms by directly integrating Newton’s equations of motion. These trajectories are crucial for understanding nanocluster melting mechanisms. Forces between atoms are calculated with an interatomic potential. Classical molecular dynamics uses models of the interatomic potential with parameters fit to some combination of experimental data and quantum mechanical calculations, such as those based on density functional theory (DFT). 

It is possible to use DFT to compute interatomic potentials at each time step in an MD simulation – sometimes referred to as Born-Oppenheimer MD (BOMD) or \textit{ab initio} MD\cite{car_unified_1985}. This method is more accurate but is considerably more computationally expensive (prohibitively expensive for clusters of tens of atoms). BOMD has been used to simulate the melting of palladium clusters\cite{luna-valenzuela_pd8_2024} and gallium clusters with changing electronic properties or competing stable solid phases\cite{wallace_melting_1997,breaux_gallium_2004,steenbergen_electronic_2012,steenbergen_geometrically_2013,steenbergen_quantum_2015,steenbergen_two_2014}.

This study uses classical MD simulation because of its accessibility and computational feasibility. Moreover, classical MD lends itself more readily to subsequent analyses involving more complex processes relevant to CNT growth (e.g., surface adsorption/desorption, diffusion, carbon dissolution, and formation of graphitic carbon). Magic numbers in Fe have been studied in small nanoclusters with both experiments\cite{sakurai_magic_1999} and spin polarization DFT simulations to capture magnetic properties \cite{kim_magic_2014,ma_structures_2007,akturk_bh-dftbdft_2016,cervantes-salguero_structure_2012,yu_theoretical_2007}. Although classical MD simulations do not capture detailed electronic magic number effects (e.g., Fe$_{7}$ and Fe$_{15}$), geometric magic numbers (e.g., Fe$_{13}$) and their properties can be extracted and may be relevant to other transition metal atoms apart from Fe.

The paper is organized as follows. Section \ref{methods} details the MD simulation method and the associated nanocluster structural and thermodynamic analysis. Section \ref{results} summarizes the results of the calculations of Fe nanocluster melting and phase transition characteristics. Section \ref{discussion} discusses the relationship between nanocluster size, structure, and melting behavior, and Section \ref{conclusion} gives concluding remarks.

\section{Methods}
\label{methods}

Molecular dynamics (MD) simulations were used to investigate cluster melting. In MD, discrete atoms are simulated in a periodic box and their motion is integrated forward in time, abiding by Newton's equations of motion.

Classical MD, which uses simple interatomic potentials, can access larger lengthscales and timescales than quantum mechanical methods. However, some long-time- and length-scale phenomena (e.g., vapor condensation, leading to nucleation and growth of large numbers of NPs) are still prohibitively expensive because MD must account for every atom's movement. Despite these limitations, MD is still useful for gaining insights into atomic-scale phenomena such as the individual nanocluster phase transitions analyzed in this work, or Gibbs free energy calculations useful for kinetic modeling of homogeneous condensation in \textit{Khrabry et al}\cite{khrabry2024gibbsfreeenergiesfe}.

Cluster melting data was obtained with classical MD simulations using the open-source LAMMPS (Large Atomic/Molecular Massively Parallel Simulator) software\cite{LAMMPS} with an embedded atom method Finnis-Sinclair (EAM-FS)\cite{finnis_simple_1984} many-body interatomic potential for Fe. The potential energy of a given atom $i$ is given by
\begin{equation}
    E_i = F_{\alpha} \left(\sum_{j \neq i} \rho_{\alpha \beta} (r_{ij}) \right) + \frac{1}{2}\sum_{j \neq i} \phi_{\alpha \beta} (r_{ij}),
\end{equation}
where $F_{\alpha}$ is the embedding energy, a function of the (modeled) electron density, $\rho_{\alpha \beta}$, contributed by neighboring atom $j$ of element $\beta$ at the site of atom $i$ of element $\alpha$. $\phi_{\alpha \beta}$ is a simple pair potential between atoms $i$ and $j$. The potential was parameterized to reproduce solid and liquid characteristics of Fe\cite{mendelev_development_2003}. Solid state binding energies agreed with those calculated using density functional theory (DFT) (supplementary material Fig.S1-3). Furthermore, the boiling point and latent heat of vaporization were validated using MD simulations of direct vapor-liquid co-existence (supplementary material Fig.S4).

Global minimum energy configurations for Fe$_n$ clusters for up to 100 atoms from Elliot et al.\cite{elliott_global_2009} (obtained with basin-hopping energy minimization) were used as starting configurations for Fe cluster structure optimization. The EAM-FS potential used in this work had a different parametrization than the that used in Ref.\cite{elliott_global_2009}. New minimum energy configurations for most of the cluster sizes were obtained using parallel tempering (PT) MD simulations, also called replica-exchange MD). In the PT simulations, 90 replicas of individual Fe$_n$ clusters were run concurrently in the canonical ensemble (NVT – constant number of atoms, N; volume, V; and temperature T) at 90 different temperatures from 250 K to 2500 K in increments of 25 K. Simulations were initiated at the global minimum energy configurations and run for 21 ns (1 ns warm-up time and 20 ns where data was collected) with a 1 fs timestep. Every 100 timesteps, each replica is proposed to swap with the replica at the temperature above it with a Metropolis acceptance probability:
\begin{equation}
    P^{acc} = min\left[1, \exp\left(\left(\frac{1}{k_B T_1} - \frac{1}{k_B T_2}\right)(E_1 - E_2)\right)\right],
\end{equation}
where $k_B$ is Boltzmann's constant, and $T_i$ and $E_i$ are the temperatures and energies of the lower ($_1$) and next higher ($_2$) temperature replica, respectively. The exchanges between replicas allow for good sampling of configurations that lie on opposing sides of an energy barrier, as each replica is able to achieve high temperatures to overcome these barriers. The replica trajectories (which traversed a wide range of temperatures) were reordered into isothermal trajectories at the 90 chosen temperatures. The lowest temperature trajectory should be a good sampling of the low-energy configurations of any given cluster size. For more than 2/3 of the cluster sizes, new lower energy configurations were obtained (after energy minimization in LAMMPS) from these low-temperature trajectories.

At lower temperatures and larger cluster sizes, as well as at temperatures of large first-order-like phase transitions, $P_{acc}$ for certain replicas can decrease to values that preclude adequate traversal between temperatures. Excluding lower temperatures (if possible), decreasing the spacing between adjacent temperatures, or employing longer simulation times can compensate for this effect. PT simulations were used in part to obtain correct low-energy configurations for subsequent calculations and observations and in part to adequately sample cluster thermodynamic space to calculate caloric curves, heat capacities, and melting temperatures. Thermodynamic data for these calculations were collected every 10 timesteps, including kinetic and potential energy. Heat capacity was calculated from the fluctuations in potential energy\cite{frantz_magic_2001,bagrets_lowering_2010}:
\begin{equation}
    \frac{C_v}{n k_B} = \frac{\langle E_{pot}^2 \rangle - \langle E_{pot} \rangle^2}{n k_B^2 T^2} + \frac{3}{2},
\end{equation}
where $\langle E_{pot}\rangle$ is the average potential energy of the cluster over the simulation time, $T$ is temperature. The $3/2$ term is added as a kinetic contribution. Melting points, $T_{C_v}$, were defined as the temperature of maximum $C_v$ for clusters with first-order-like phase transitions, and the temperature of steepest $C_v$ slope for clusters with second-order-like phase transitions.

Simple single-cluster isothermal MD simulations were also run for further mechanistic melting analysis. The simulations were initiated from the global minimum energy configurations obtained from the PT simulations and run in the NVT ensemble for 20 ns at 75 temperatures from 50 K to 3750 K in increments of 50 K, with atomic configurations collected every 1 ps. For all PT and simple NVT simulations in this work, the canonical stochastic Langevin dynamics (CSLD) thermostat\cite{Accurate_sampling_Bussi} was invoked every 100 timesteps. Atomic Lindemann indices:
\begin{equation}
    \delta_i = \frac{1}{n - 1} \sum_{i \neq j} \frac{\sqrt{\langle r_{ij}^2 \rangle - \langle r_{ij} \rangle^2}}{\langle r_{ij} \rangle},
\end{equation}
and whole cluster Lindemann index:
\begin{equation}
    \delta_c = \frac{2}{n(n-1)} \sum_{i > j} \frac{\sqrt{\langle r_{ij}^2 \rangle - \langle r_{ij} \rangle^2}}{\langle r_{ij} \rangle},
\end{equation}
were calculated at each temperature to measure increases in atomic motion due to melting. A time average is represented by $\langle\rangle$ and $r_{ij}$ is the pairwise distance between atoms of indices $i$ and $j$. The atomic distance from the cluster center of mass (CoM), $r_{i,CoM}$, was also calculated for each temperature to yield structural information about core melting.

\section{Results}
\label{results}
Molecular dynamics simulations were performed for individual clusters with up to 100 atoms at temperatures spanning the melting transition. The definition of melting is ambiguous for nanoclusters, and melting can occur over a wide range of temperatures, with the surface often melting before the core. Moreover, due to the presence of many configurations that are close in energy and the dynamic-coexistence (isomerization) nature of nanocluster melting\cite{rice_evolution_2009,schebarchov_solid-liquid_2006,aguado_neither_2016}, a single unambiguous melting point temperature does not adequately describe the transition. This work attempts to demystify Fe cluster melting by analyzing cluster energetics and mechanisms of melting with cluster structure in mind.

\subsection{Structure evolution with cluster size}

\begin{figure}
\centering
    \includegraphics[width=0.9\linewidth]{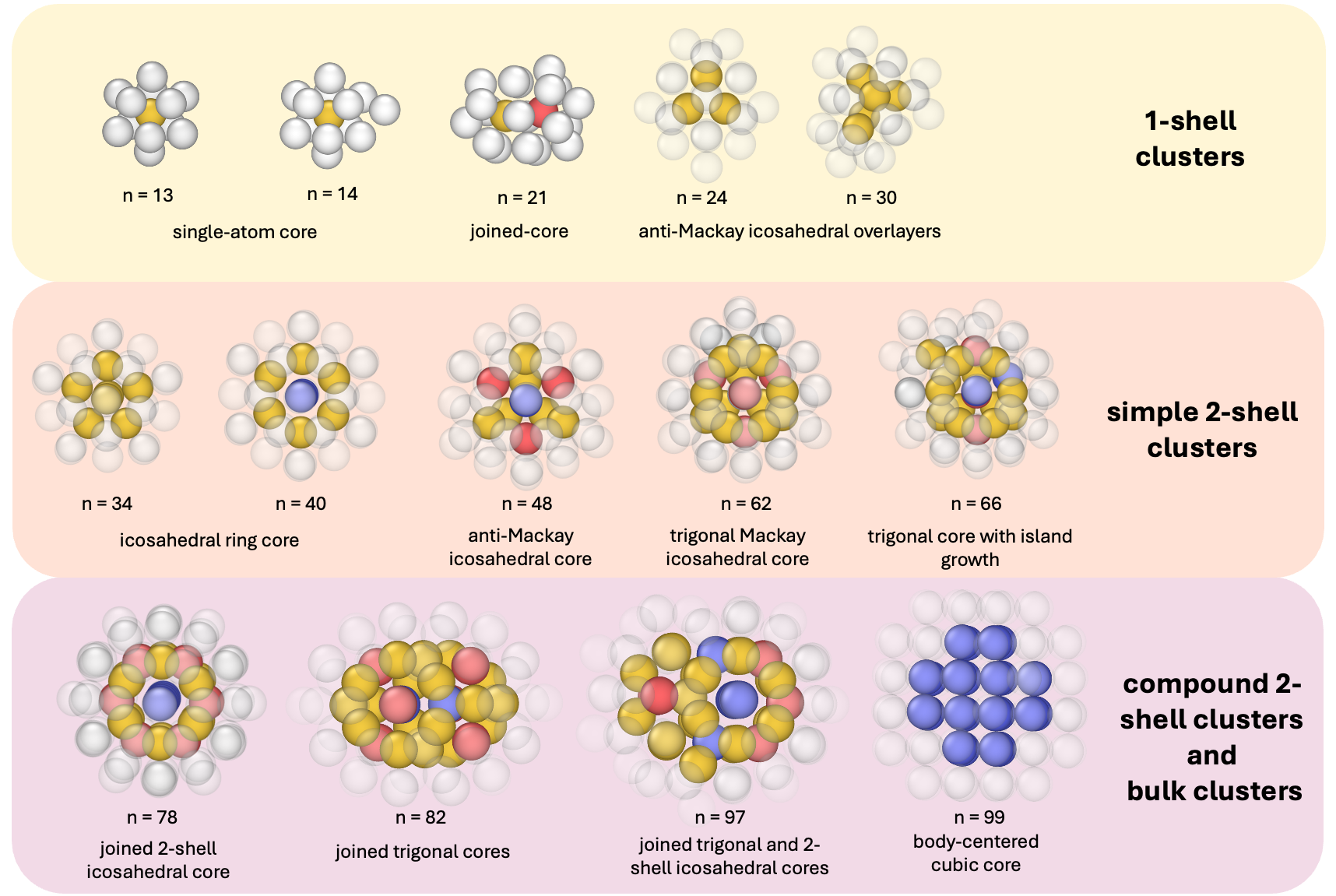}
\caption{Summary snapshots of global minimum energy structures obtained from parallel tempering simulations for 1-shell, simple 2-shell, and compound 2-shell and bulk structure clusters. Surface atoms in many snapshots are made transparent, core atoms are colored by structure type: icosahedral (yellow), hexagonal close-packed (red), or body-centered-cubic (purple). Structure identification coloring is used to guide the eye and should not be interpreted strongly.}
\label{fig: min_structs}
\end{figure}

As clusters increase in size,  the structures of minimum energy configurations do not include the simple addition of 1 atom at a time to successive shells with similar cluster structures. Instead, several transitions in structure occur with increasing size. Fig.\ref{fig: min_structs} summarizes the evolution of the most stable cluster core structures determined from parallel tempering simulations.
Undercoordinated surface atoms are made transparent to ease the viewing of internal core structures. Polyhedral template matching in Ovito\cite{ovito} is used to help identify core atom structure with an RMSD of 0.32, a lenient threshold. Yellow, red, and purple atoms correspond to icosahedral (ico), hexagonal-close packed (hcp), and body-centered cubic (bcc) structures, respectively. While the structure assignments are loose, they give an idea of the relative coordination of an atom compared to a normal icosahedral structure: red atoms are overcoordinated (have more bonding partners), and purple atoms are even more so. Structures with bcc atoms at the edge of the cluster core tended to undergo structural shifts in larger subsequent cluster sizes.

\textbf{1-shell clusters} (top row of Fig.\ref{fig: min_structs}) start out with a single atom in the core surrounded by surface atoms ($Fe_{13}$). As more atoms are added, they are placed on the surface alone and undercoordinated (with dangling bonds—like $Fe_{14}$) or incorporated into an expanded surface structure $Fe_{15-17}$. Eventually, enough atoms are present to subsume another atom into the core in a joined-core structure ($Fe_{18-22}$). Atoms are added into the core one at a time, forming small icosahedra ($Fe_{26-30}$) and filling out the overlayers until the cluster subsumes another atom into its core.

\textbf{Simple 2-shell clusters} (middle row of Fig.\ref{fig: min_structs}) begin with a small planar ring and one inner core atom $Fe_{32}$ but only become stable with 2 inner core atoms surrounded by 5 inner-ring atoms ($Fe_{34}$). There is competition with icosahedral 1-shell structures for many sizes. Extra atoms are added to the surface until the ring can be expanded to 6 atoms ($Fe_{40}$). Non-planar core structures compete for stability with 6-atom inner-ring core structures like the anti-Mackay icosahedral $Fe_{48}$ cluster. More atoms are added until the 6-atom ring-shell structure forms a Mackay-icosahedral trigonal-symmetric core ($Fe_{62}$), which remains stable for many cluster sizes, accumulating surface island atoms atom the same core structure (e.g., $Fe_{66}$).

\textbf{Compound 2-shell clusters} (bottom row of Fig.\ref{fig: min_structs}) become stable with 76 atoms, where 2 axially connected 2-shell icosahedra form a tunnel-like structure ($Fe_{78}$). With a few more atoms, this structure competes for stability with interpenetrating trigonal-symmetric Mackay-icosahedral cores ($Fe_{82}$), then forms hybrid structures with 1 trigonal-symmetric core attached to a double-2-shell icosahedron core ($Fe_{97}$). These hybrid structures compete for stability with bulk-type bcc clusters ($Fe_{99}$). It is unclear if 3-shell icosahedral-type clusters are possibly preferable to the bcc clusters at larger sizes.

\subsection{Cluster energetics}
\begin{figure}
    \centering
    \includegraphics[width=0.8\textwidth]{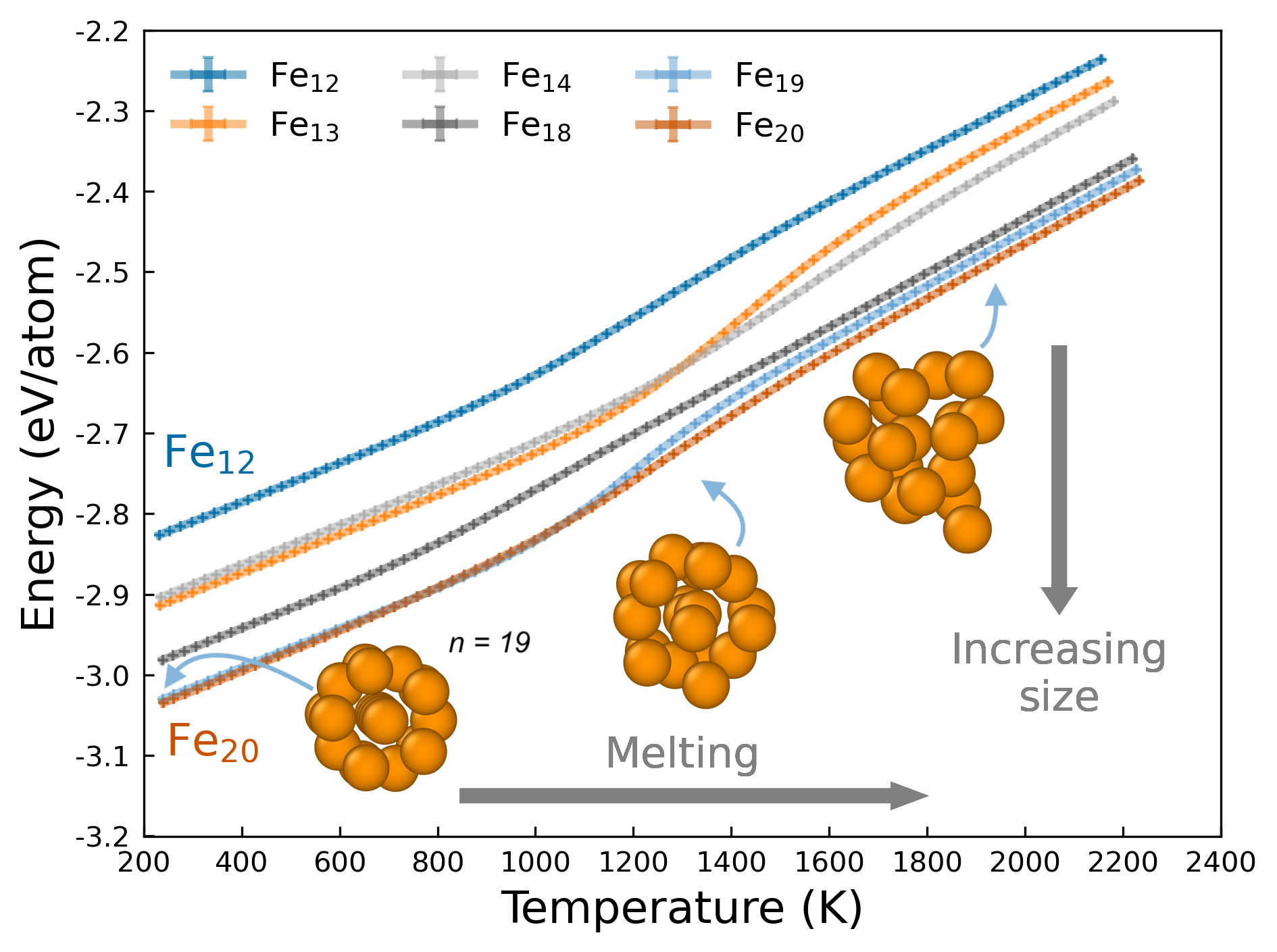}
    \caption{MD caloric curves for selected clusters from Fe$_{12}$-Fe$_{20}$. Each point corresponds to an isothermal trajectory from parallel tempering simulations. The closed-shell clusters ($Fe_{13}$ in orange and $Fe_{19}$ in light blue) have larger rises in energy in their transition from solid to liquid phase than their neighboring sizes due to comparably large latent heat of melting.}
    \label{fig: series_caloric_curves}
\end{figure}

To quantify the melting behavior of iron clusters, caloric curves of total energy (eV/atom) vs. temperature (K) were obtained for each cluster size. Each data point corresponds to a stitched-together (reordered) parallel tempering (PT) MD trajectory at one cluster size and temperature (Fig.\ref{fig: series_caloric_curves}). Temperature was determined from time-averaged cluster kinetic energies:
\begin{equation}
    T_{MD} = \frac{2 E_{kin}}{n_{DOF} k_B},
\end{equation}
where $n_{DOF}$ is the number of kinetic degrees of freedom of the cluster, equal to $3n$ (not $3n-6$, which would apply to rotationally and translationally fixed nonlinear clusters). The CSLD thermostat activates all degrees of freedom of the clusters, including translation and rotation.

Caloric curves qualitatively differ by cluster size – a phenomenon that this work attempts to address. For instance, \textit{closed-shell} clusters (e.g., Fe$_{13}$ and Fe$_{19}$ in Fig.\ref{fig: series_caloric_curves} – many have perfect or interpenetrating icosahedral structures) exhibit phase transitions with large increases in total energy upon melting (latent heat) relative to neighboring clusters. Adding an atom to a closed-shell cluster causes minor differences in solid-state energy, but removing an atom from a closed-shell cluster drastically increases the cluster's energy.

For liquid clusters (right side of Fig.\ref{fig: series_caloric_curves}), specific energy decreases with cluster size, $n$: $E_{n-1} > E_n > E_{n+1}$. A cluster with a higher $n$ has more bonds per atom due to a smaller surface-to-volume ratio, leading to a lower specific energy. Accordingly, the \textit{difference} in specific energy between adjacent sizes (spacing between curves) generally decreases with increasing $n$, as the difference in surface-to-volume ratios decreases with greater $n$. The trend is violated in the solid phase regime (left) for closed-shell magic number clusters, whose energy curves then cross their $n$+1 neighbor cluster ($Fe_{14}$ and $Fe_{20}$ in Fig.\ref{fig: series_caloric_curves}) upon phase transition into the liquid phase regime (right), restoring the trend.

\begin{figure}
\centering
\begin{minipage}{.45\textwidth}
  \centering
  \includegraphics[width=\linewidth]{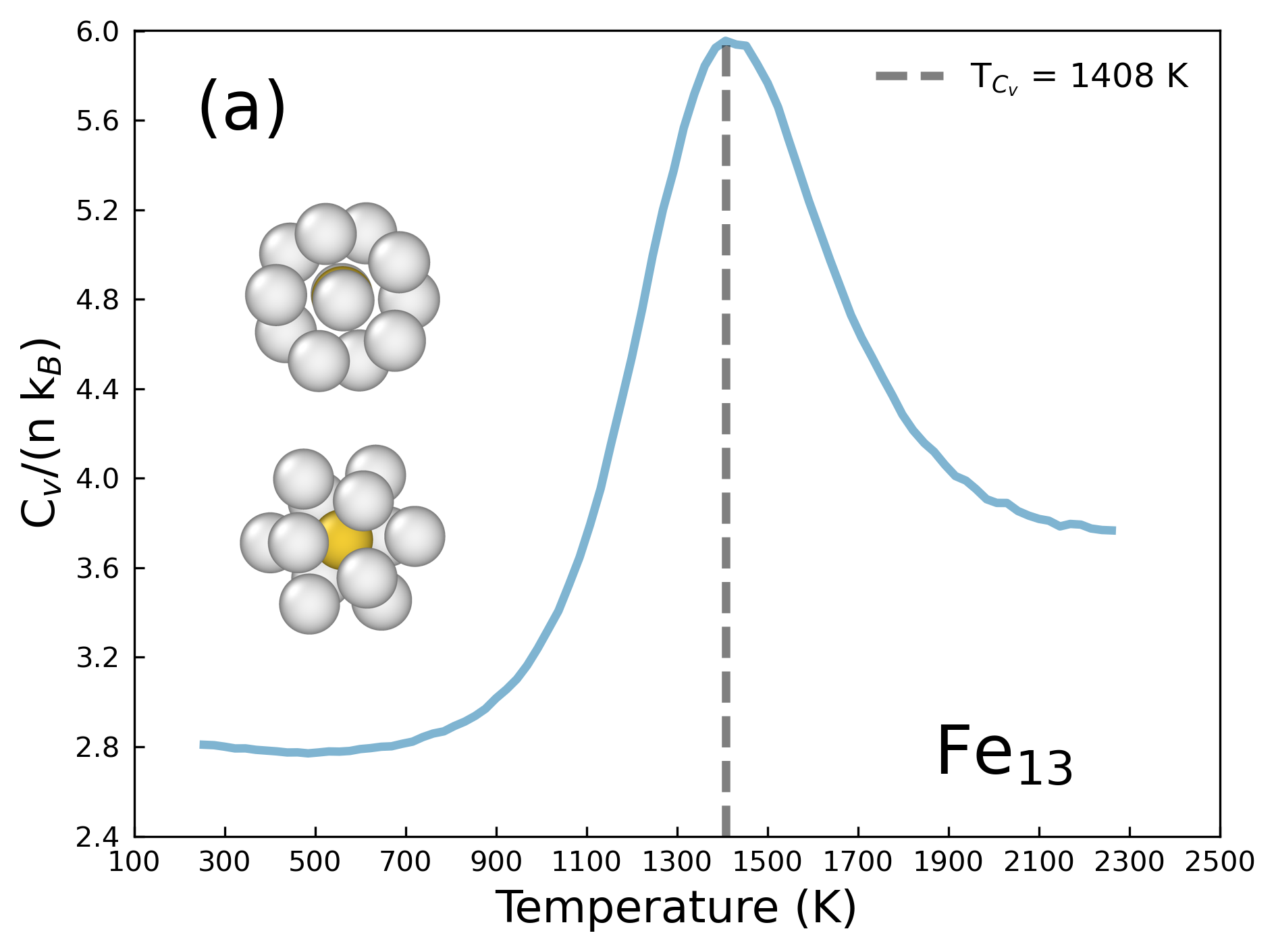}
\end{minipage}
\begin{minipage}{.45\textwidth}
  \centering
  \includegraphics[width=\linewidth]{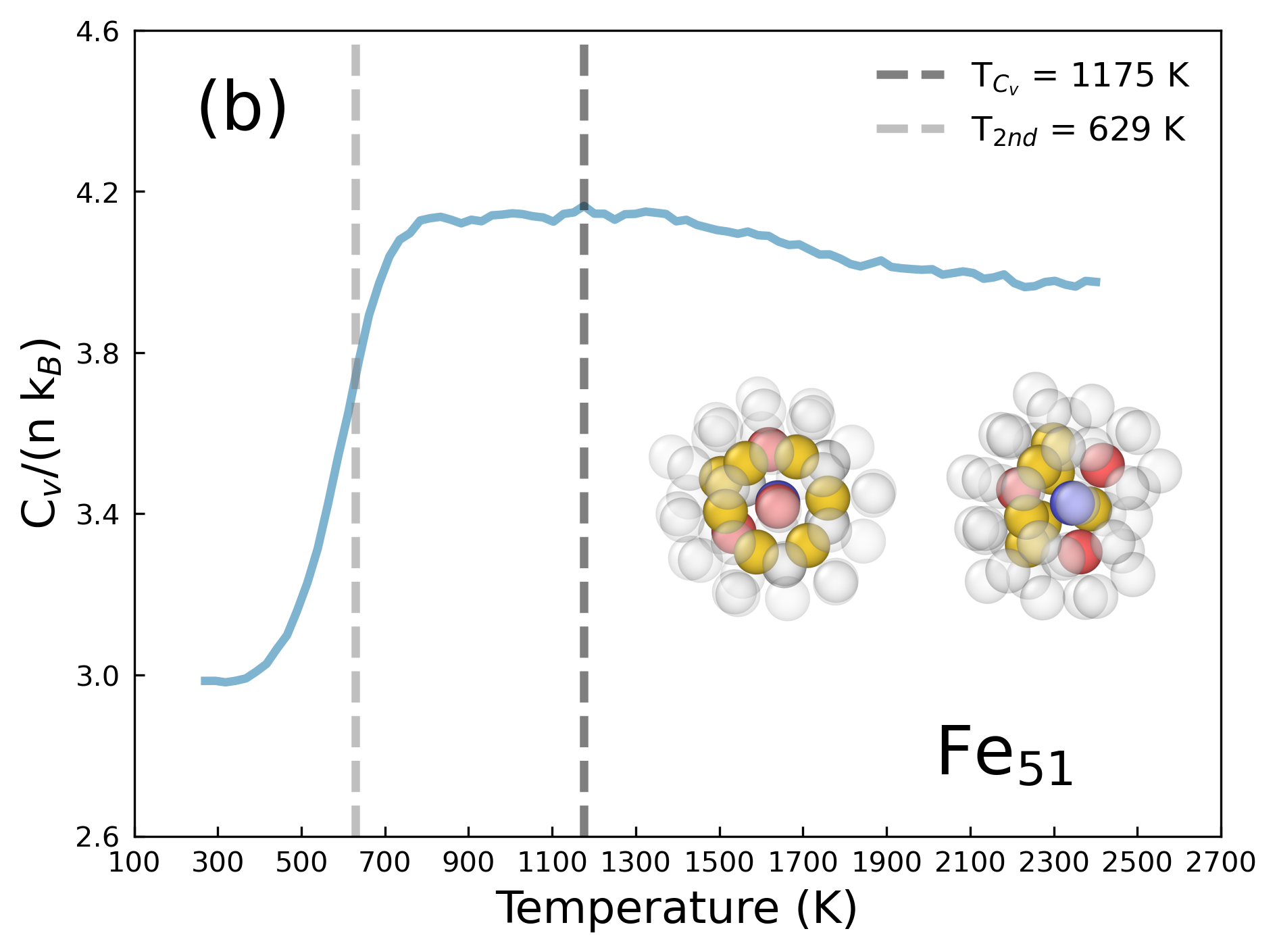}
\end{minipage}
\begin{minipage}{.45\textwidth}
  \centering
  \includegraphics[width=\linewidth]{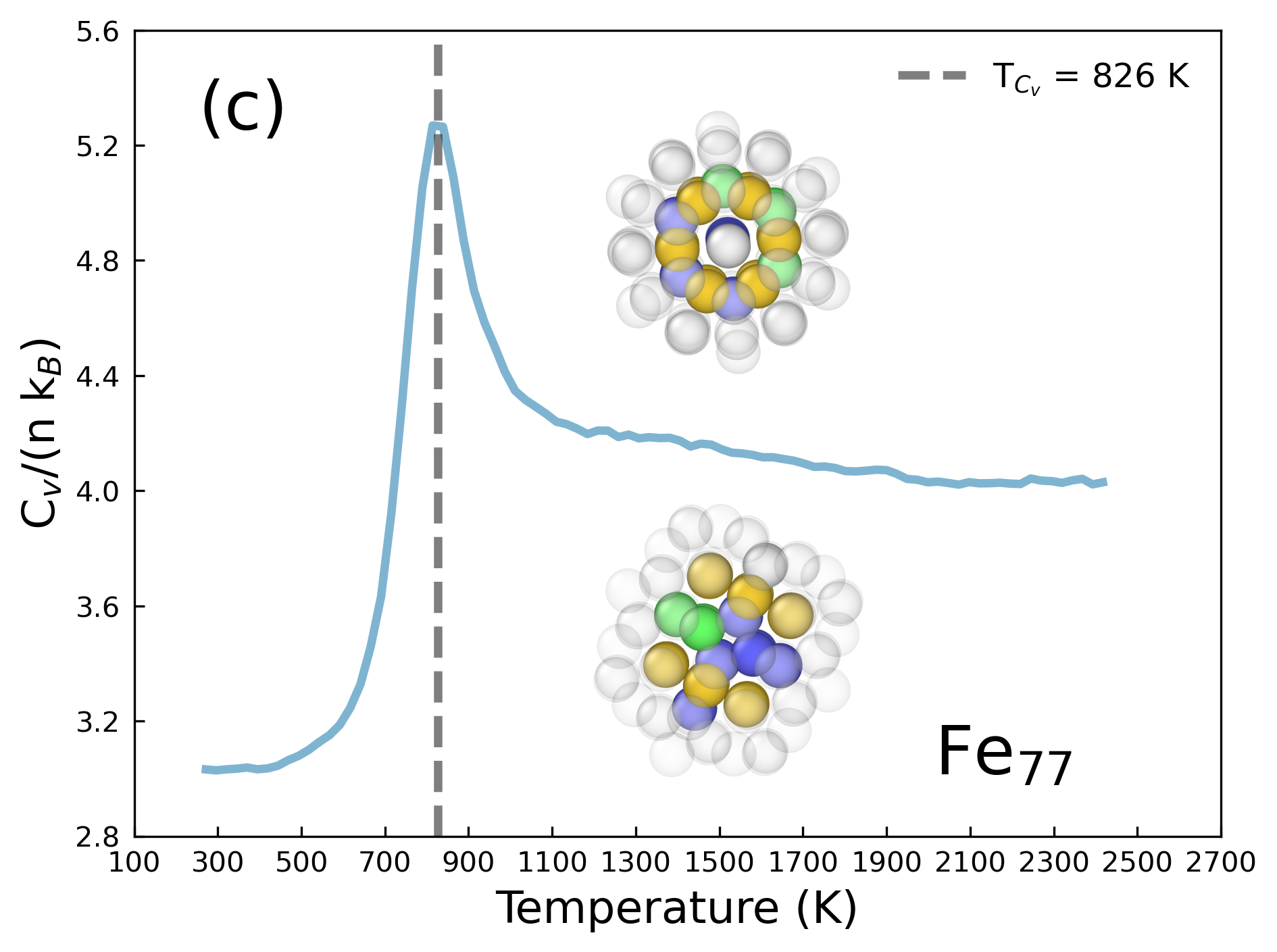}
\end{minipage}
\caption{Heat capacity curves for 3 selected cluster sizes: \textbf{(a)} Closed-shell $Fe_{13}$ has a first-order-like phase transition with a tall and wide peak at a high melting point. \textbf{(b)} $Fe_{51}$ has a second-order-like phase transition with no pronounced peak in its heat capacity curve. \textbf{(c)} Near-closed $Fe_{77}$ also has a first-order-like transition but with a narrower peak and a lower melting point. Undercoordinated surface atoms are transparent. Core atoms are colored with polyhedral template matching: yellow, red, purple, and green atoms correspond to icosahedral (ico), hexagonal-close packed (hcp), and body-centered cubic (bcc) structures, and face-centered cubic (fcc), respectively.}
\label{fig: individual_Cvs_fig}
\end{figure}

Heat capacity curves were calculated from the PT simulations for each cluster size to aid in melting point determination. Individual $C_v$ curves normalized by $k_B$ and $n$ are given for closed-shell cluster $Fe_{13}$, $Fe_{51}$, and nearly closed-shell cluster $Fe_{77}$ in Figs.\ref{fig: individual_Cvs_fig}a-c. The $C_v$ curves visually vary more starkly than the caloric curves, and this work attempts to investigate these differences.
Most cluster sizes have $C_v$ curves with notable peaks corresponding to the latent heat present in first-order-like phase transitions. Within the size range studied, larger clusters tended to have sharper $C_v$ peaks at lower temperatures than smaller clusters. All closed-shells and most nearly closed-shell clusters (e.g. $Fe_{13}$ and $Fe_{77}$ in Fig.\ref{fig: individual_Cvs_fig}a,c) exhibit first-order-like phase-transition peaks in their $C_v$ curves. The 'energetic' melting temperature, $T_{C_v}$, is defined as the temperature of maximum heat capacity. This maximum heat capacity, max$(C_v)$, can be thought of as a proxy for the latent heat of melting for the cluster, which will be discussed later on.

Twenty-eight cluster sizes in the range studied do not show a pronounced peak in $C_v$ (e.g., $Fe_{51}$ in Fig.\ref{fig: individual_Cvs_fig}c), having either a very small or non-existent latent heat of melting, indicating a second-order-like phase transition. For these second-order-like melting clusters, a 2nd $T_{C_v}$ is defined as the temperature of the maximum steepest ascent in $C_v$, labeled $T_{2nd}$. This temperature corresponds to a maximum in the second derivative of the caloric curve $\left(\frac{\partial^2 E}{\partial T^2}\right)_v$, the expected temperature of a second-order phase transition. Heat capacity curves for all cluster sizes from 10-100 atoms (with $T_{C_v}$ and $T_{2nd}$ graphically labeled) can be found in the Cv-plots folder of the SI.

\begin{figure}
\centering
\begin{minipage}{.45\textwidth}
  \centering
  \includegraphics[width=\linewidth]{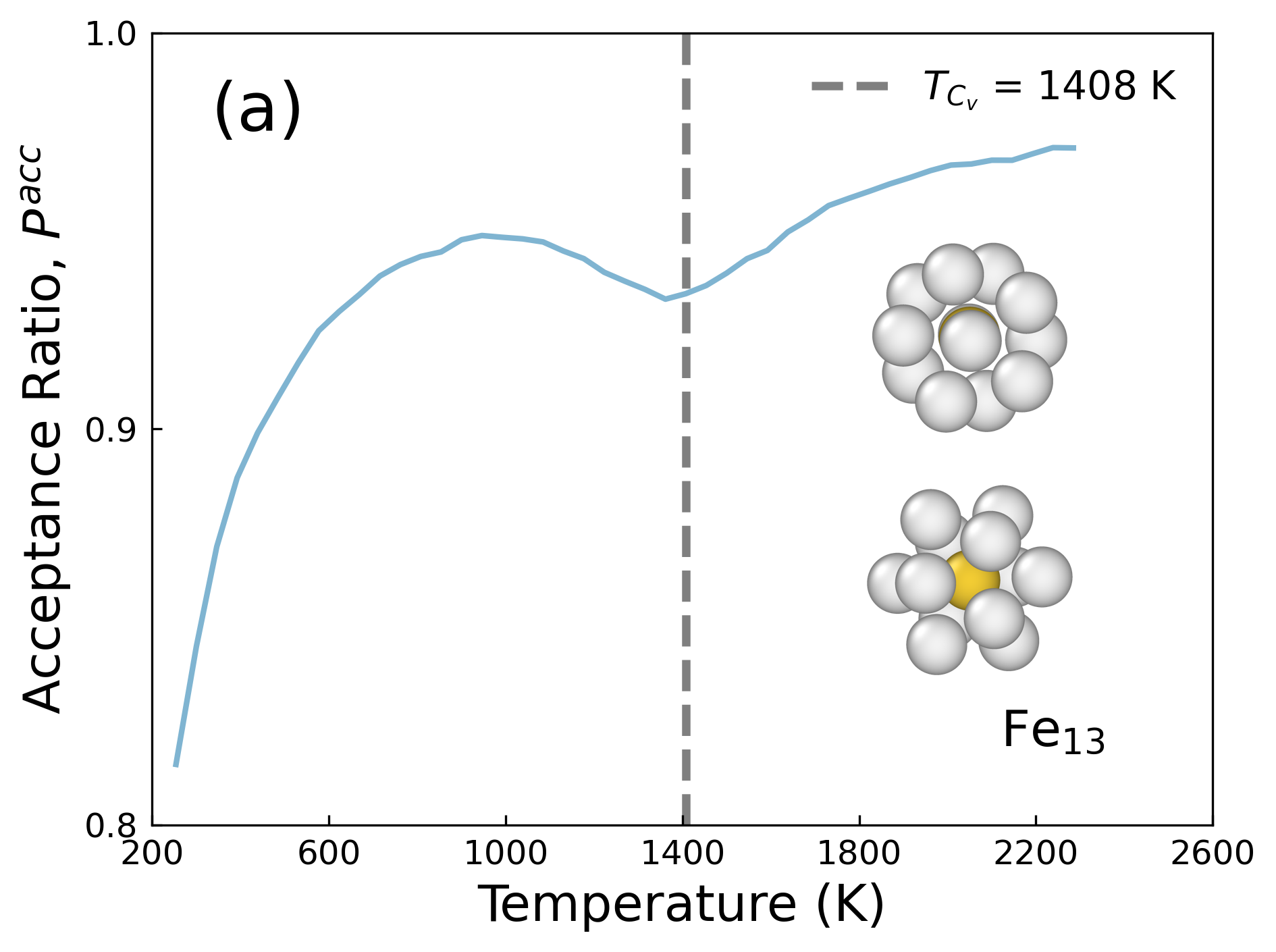}
\end{minipage}
\begin{minipage}{.45\textwidth}
  \centering
  \includegraphics[width=\linewidth]{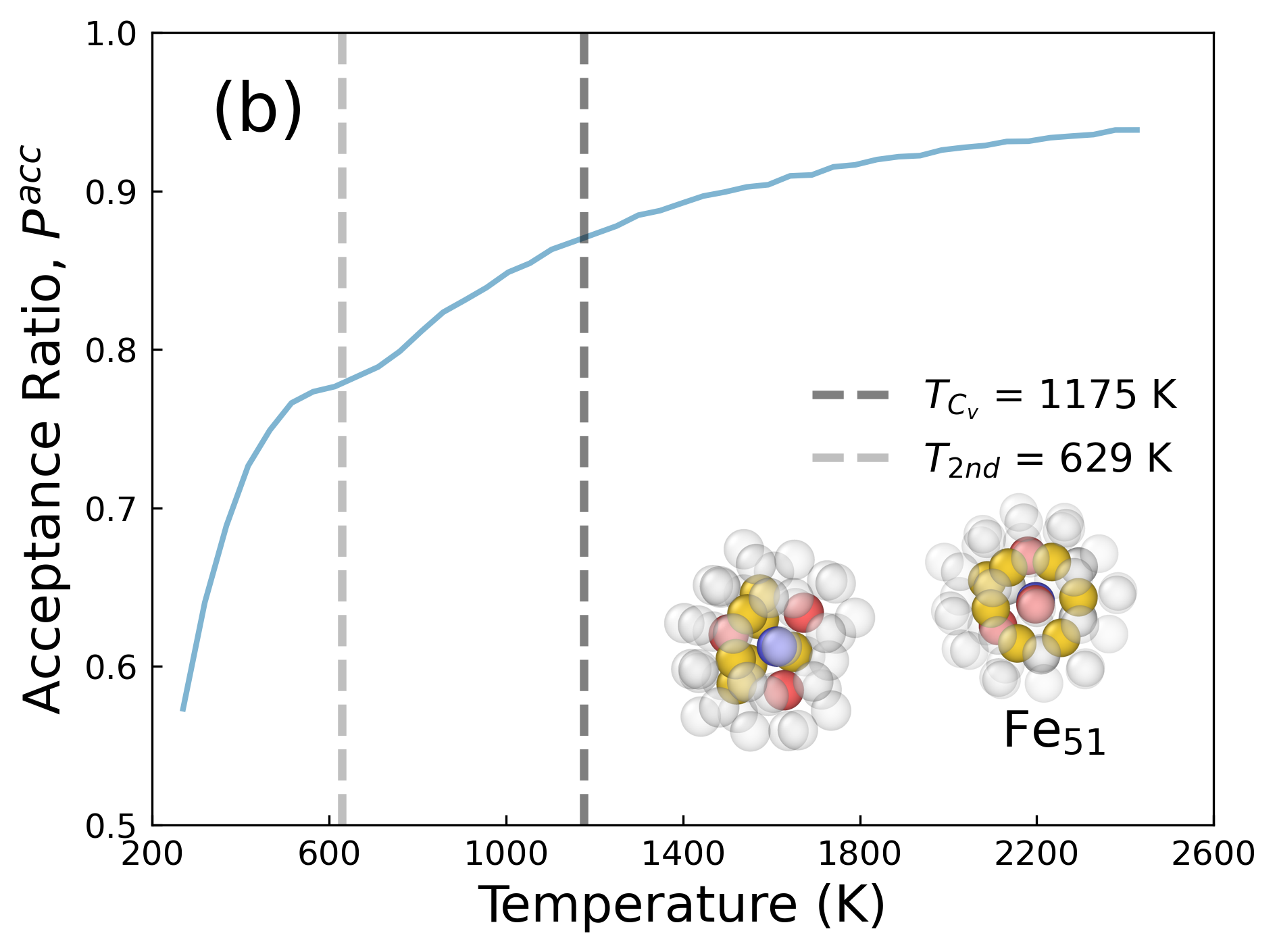}
\end{minipage}
\begin{minipage}{.45\textwidth}
  \centering
  \includegraphics[width=\linewidth]{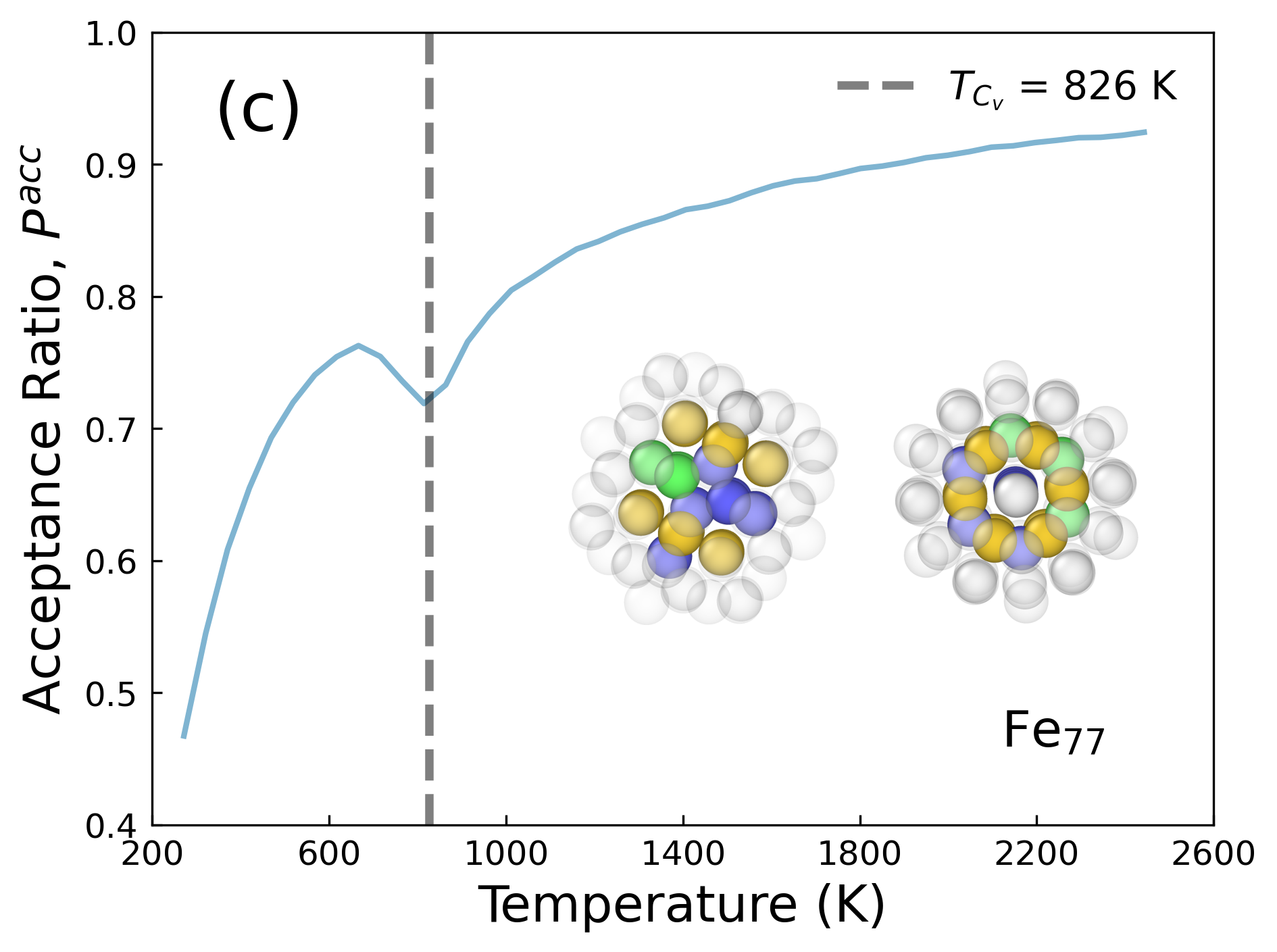}
\end{minipage}
\caption{Acceptance ratio curves during parallel tempering MD simulations of \textbf{(a)} closed-shell $Fe_{13}$ and \textbf{(b)} far-from-closed $Fe_{51}$ which features a dent in its $P^{acc}$ at $T_{C_v}$ defined from its heat capacity curve – and \textbf{(c)} near-closed $Fe_{77}$. $Fe_{13}$ and $Fe_{77}$ both show large $P^{acc}$ dips at their phase transition temperatures.}
\label{fig: individual_Pacc_fig}
\end{figure}

At the temperatures near a phase transition region in a PT simulation with uniform temperature spacing, the acceptance probability, $P^{acc}$, of parallel tempering replica swaps is reduced for isomerizing clusters due to less overlap of energy distributions between temperature-adjacent replicas (assuming a constant increment of temperatures between replicas). Acceptance ratio curves for PT simulations are given for $Fe_{13}$, $Fe_{51}$, and $Fe_{77}$ in Fig.\ref{fig: individual_Pacc_fig}a-c, and for all clusters in the Pacc-plots folder of the SI.

In cluster sizes exhibiting first-order-like phase transitions, $P_{acc}$ is reduced around $T_{C_v}$, where rapid cluster isomerization occurs. For cluster sizes exhibiting second-order-like phase transitions (e.g., $Fe_{51}$ in Fig.\ref{fig: individual_Cvs_fig}b), the $P_{acc}$ reduction occurs near $T_{2nd}$ (629 K for $Fe_{51}$),  not $T_{C_v}$ (1175 K for $Fe_{51}$), due to the low-temperature onset of interchange between structural isomers. Therefore, for second-order-like cluster sizes, $T_{C_v}$ lacks meaning, as it does not indicate any part of melting.

\begin{figure}
\centering
\begin{minipage}{.45\textwidth}
  \centering
  \includegraphics[width=\linewidth]{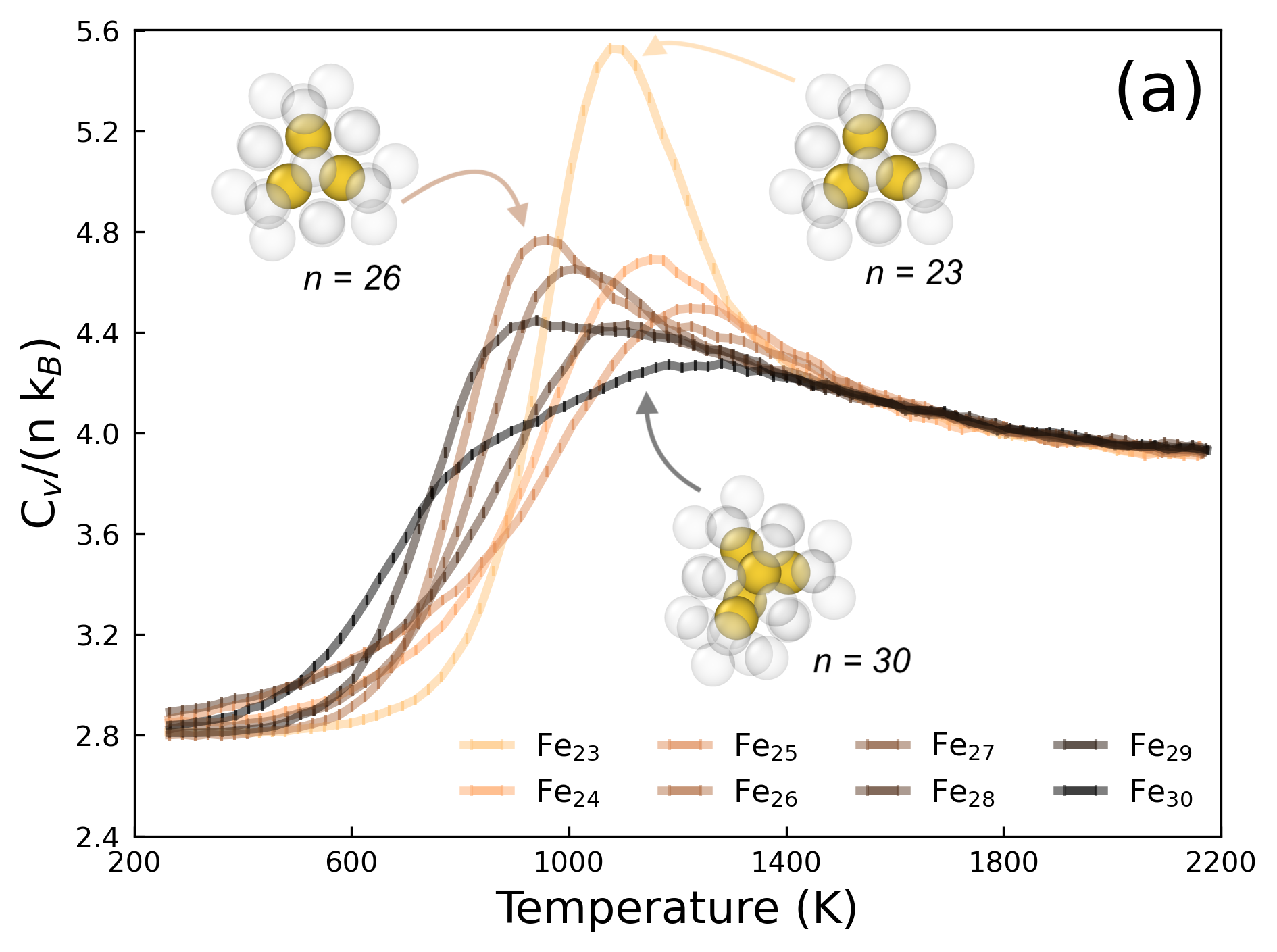}
\end{minipage}
\begin{minipage}{.45\textwidth}
  \centering
  \includegraphics[width=\linewidth]{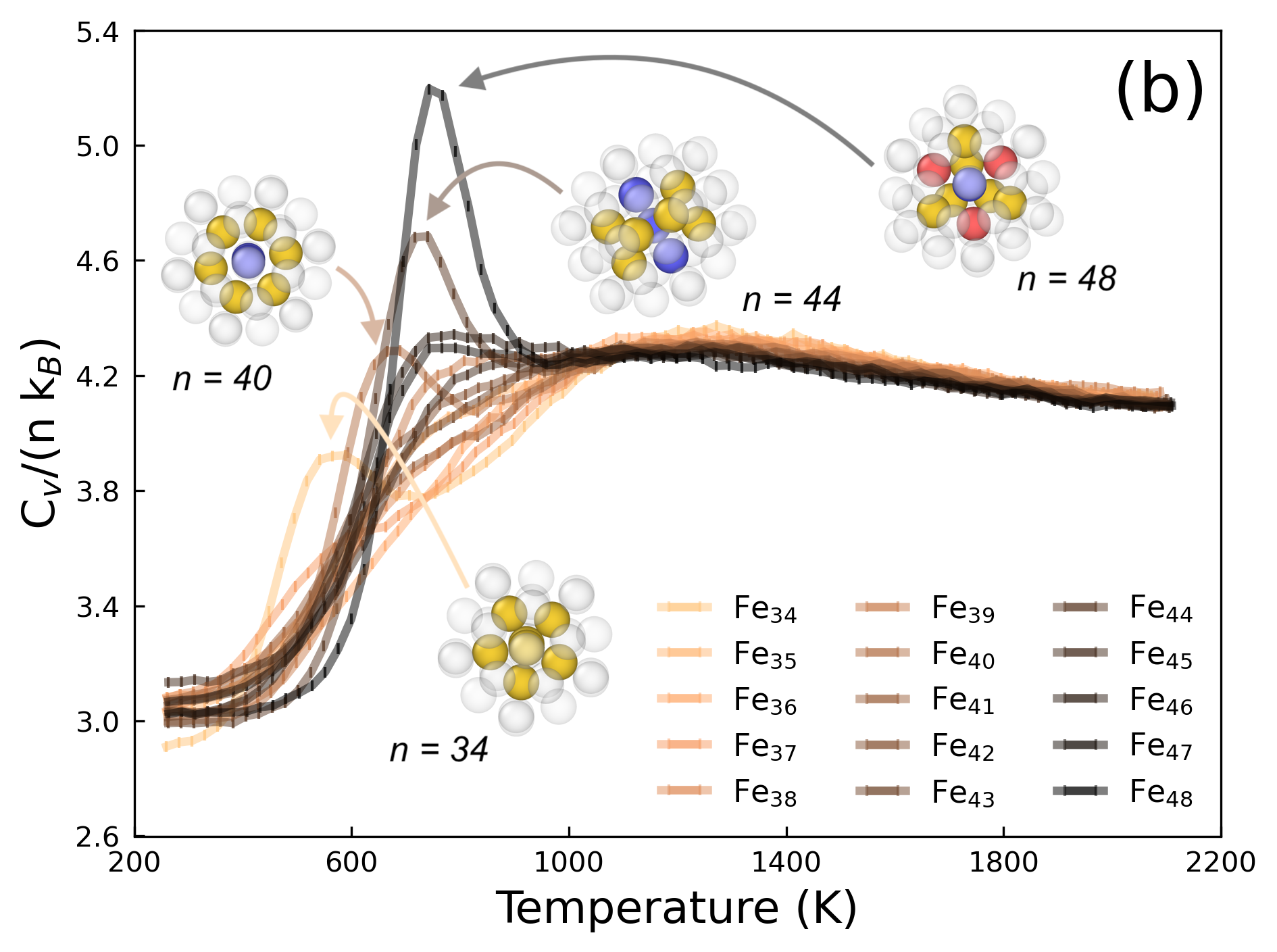}
\end{minipage}
\begin{minipage}{.45\textwidth}
  \centering
  \includegraphics[width=\linewidth]{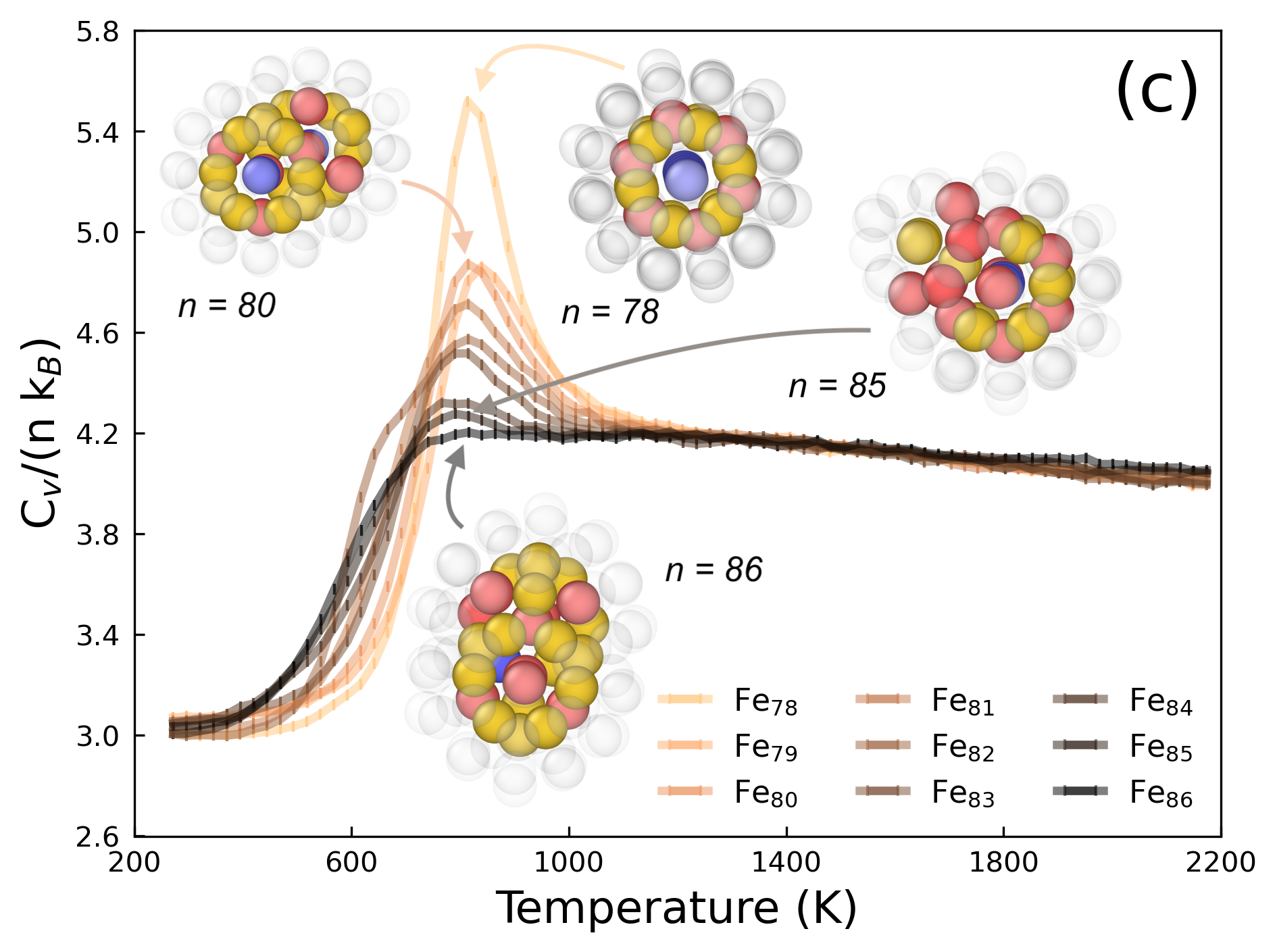}
\end{minipage}
\caption{Selected series of consecutive $C_v$ curves including clusters with first- and second-order-like phase transitions for \textbf{(a)} $Fe_{23}$ - $Fe_{30}$ (structure snapshots for closed-shell $Fe_{23}$, $Fe_{26}$ and far-from-closed $Fe_{30}$), \textbf{(b)} $Fe_{33}$ - $Fe_{48}$ (structure snapshots for closed-shell $Fe_{48}$ and $Fe_{44}$), and \textbf{(c)} $Fe_{78}$ - $Fe_{86}$ (structure snapshots for closed-shell $Fe_{78}$ and far-from-closed $Fe_{86}$).}
\label{fig: series_Cvs_fig}
\end{figure}

To show the evolution  of $C_v$ curves across small, medium, and large cluster sizes, multiple series of $C_v$ curves are plotted for different size ranges in Figs.\ref{fig: series_Cvs_fig}a-c. While adjacent $C_v$ curves can differ from each other at low temperatures, curves of adjacent cluster sizes converge at higher temperatures once all clusters are completely melted. The temperature of convergence upon melting is largest for smaller clusters (near 1600 K in Fig.\ref{fig: series_Cvs_fig}a), and decreases at larger cluster sizes to around 1200 K in Fig.\ref{fig: series_Cvs_fig}b-c.
For single-shell clusters ($10 \leq n \leq 31$), highly stable closed-shell structures reside only a few atoms away from other closed-shell cluster sizes. Clusters with 13, 19, 23, and 26 atoms are all closed-shell sizes, making most single-shell clusters nearly closed-shell and first-order-like in melting transition (with the exception of $Fe_{17}$ and $Fe_{22}$). Small 2-shell cluster sizes ($31 \leq n \leq 50$), predominantly have second-order-like phase transitions, except for a few clusters with closed shells, which have first-order-like phase transitions: $Fe_{34}$, $Fe_{40}$, $Fe_{44}$, and $Fe_{48}$. At large cluster sizes ($n \geq 51$), closed-shell clusters are generally sparse, with many clusters before and after having smaller first-order-like $C_v$ peaks at similar temperatures (Fig.\ref{fig: series_Cvs_fig}c).

In the $C_v$ curves of $Fe_{34}$ and $Fe_{40}$, the first-order peaks occur at low temperatures  (Fig.\ref{fig: series_Cvs_fig}b), and peak at a maximum $C_v$ lower than the liquid phase $C_v$. $T_{C_v}$ for $Fe_{34}$ and $Fe_{40}$ were calculated using the above definition, but $T_{2nd}$ is defined as the temperature of their first-order peak's maximum. These weak, low-temperature, first-order peaks in $C_v$ were noted in \textit{Frantz (2001)} in Lennard-Jones Ar\cite{frantz_magic_2001} for many cluster sizes between 30 and 40. Their appearance was attributed to a restructuring in the minimum energy configurations from anti-Mackay to Mackay icosahedral overlayers. They also observed shallower and broader primary $C_v$ peaks that appear to be second-order-like (apart from this weak, narrow, low-temperature, first-order peak). While second-order phase transitions are not mentioned, they reported that the $C_v$ peak-broadening coincided with cluster sizes with high densities of configurations with close-lying energies.

It should be noted that the minimum solid cluster heat capacity can fall below $3 n k_B$, especially for small cluster sizes. The energy-absorbing degrees of freedom for nonlinear clusters with $n$ atoms is given by\cite{smith2008quantum}:
\begin{equation}
    f = n_{translation} + n_{rotation} + 2(n_{deformation}) = 3 + 3 + 2(3n - 6) = 6n - 6,
\end{equation}
where $n_{translation}$, $n_{rotation}$, and $n_{deformation}$ are the number of modes of translation, rotation, and deformation, respectively. Therefore, the minimum normalized heat capacity is:
\begin{equation}
\label{eq: minCv}
    \frac{C_{v,min}}{nk_B} = \frac{f}{2n} = \frac{6n - 6}{2n} = 3 - \frac{3}{n}.
\end{equation}
As equation \ref{eq: minCv} suggests, the minimum heat capacity per atom for clusters asymptotically approaches $3$ with increasing size.

As mentioned, the max$(C_v)$ can be a proxy for the latent heat of melting for the cluster. In an ideal case, one would integrate $C_v$ over the cluster's melting range to obtain the value of the latent heat; however, it is impossible to systematically identify a cluster's melting range, as $C_v$ before and after melting are generally not constants. It is also likely that the melting ranges of different cluster sizes vary strongly, making max$(C_v)$ only a qualitative indicator of a cluster's latent heat of melting. Nonetheless, a low max$(C_v)$ could be a litmus test for clusters with second-order-like phase transitions.

\begin{figure}
    \centering
    \includegraphics[width=0.7\linewidth]{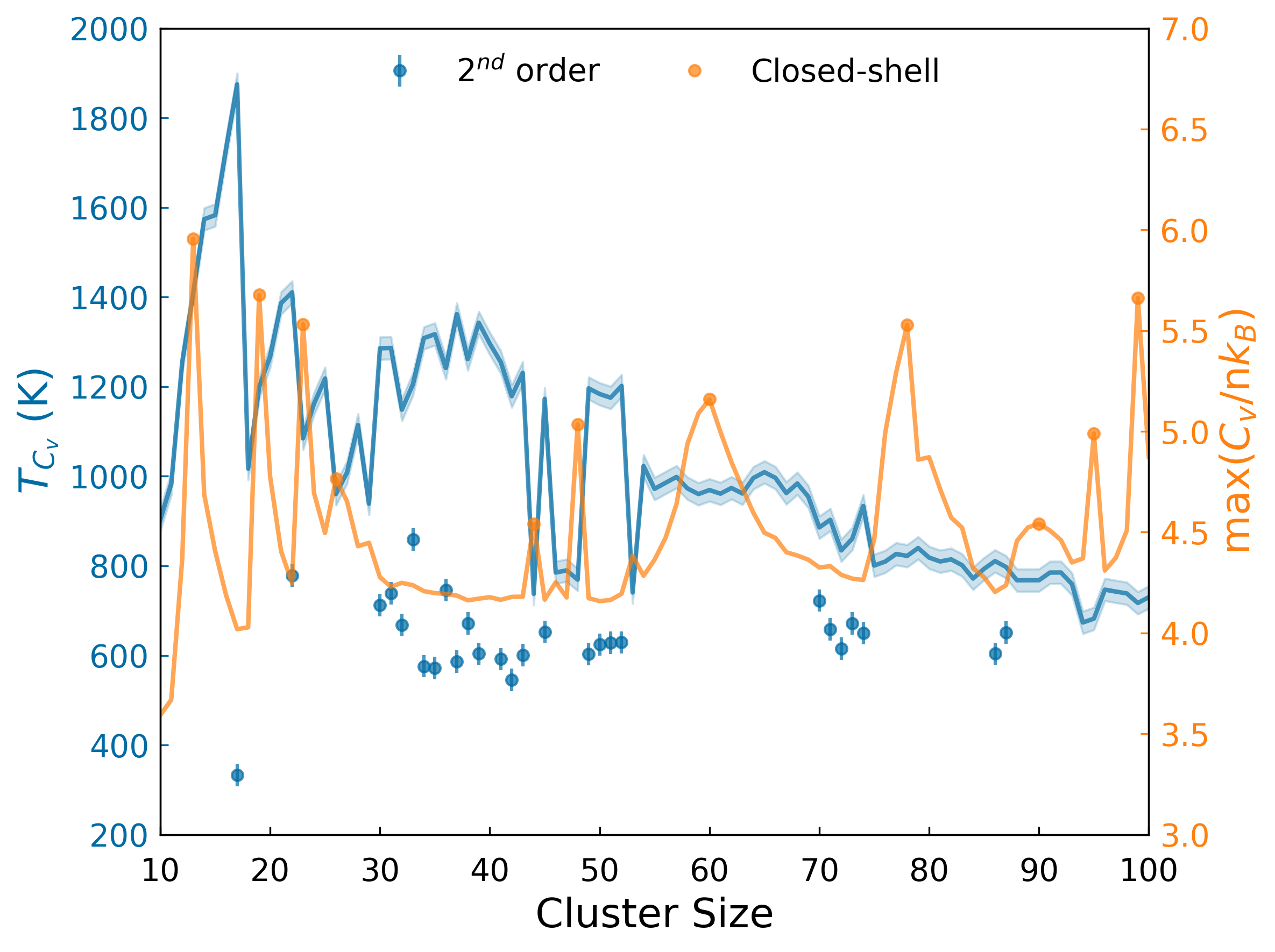}
\caption{Dual plot of $T_{C_v}$ (blue) and max$(C_v/nk_B)$ (orange) as a function of cluster size. Clusters with second-order-like melting reside at many sizes where max($C_v$) is small.}
\label{fig: max_Cvs_fig}
\end{figure}

The evolution of the strength of the first-order peaks can also be monitored with max$(C_v)$, as closed-shell clusters have higher peaks than their neighboring sizes, which generally have higher maxima than clusters with second-order-like melting. Fig.\ref{fig: max_Cvs_fig} compares $T_{C_v}$ and max$(C_v)$ as a function of cluster size, with closed-shell cluster size (orange points) and sizes with second-order-like transitions (blue points) indicated.

A high maximum in $C_v$ (i.e. proximity to a closed-shell structure) has a relatively weak effect on $T_{C_v}$; many smaller closed-shell sizes have lower melting points than the few cluster sizes that follow (e.g., $T_{C_v}$ for $Fe_{14}$-$Fe_{17}$ are higher than that of $Fe_{13}$). It is worth noting that many of the cluster sizes with the lowest max$(C_v)$ have second-order-like phase transitions. Among adjacent clusters within the peaks in max$(C_v)$ (between 50 and 86 atoms), clusters tend to build upon a basic structure (e.g., a trigonal-symmetric 2-shell structure) until a transition to a more accommodating structure is available (like the joined 2-shell icosahedra in $Fe_{78}$ – this structure first becomes preferable in $Fe_{75}$ at the left of the max$(C_v)$ peak). The spaces between the max($C_v$) peaks indicate the potential stability of multiple core configurations, in many cases inducing second-order-like melting. The interplay between cluster structure and melting properties is discussed in the rest of the work.

\begin{figure}
\centering
\includegraphics[width=0.9\linewidth]{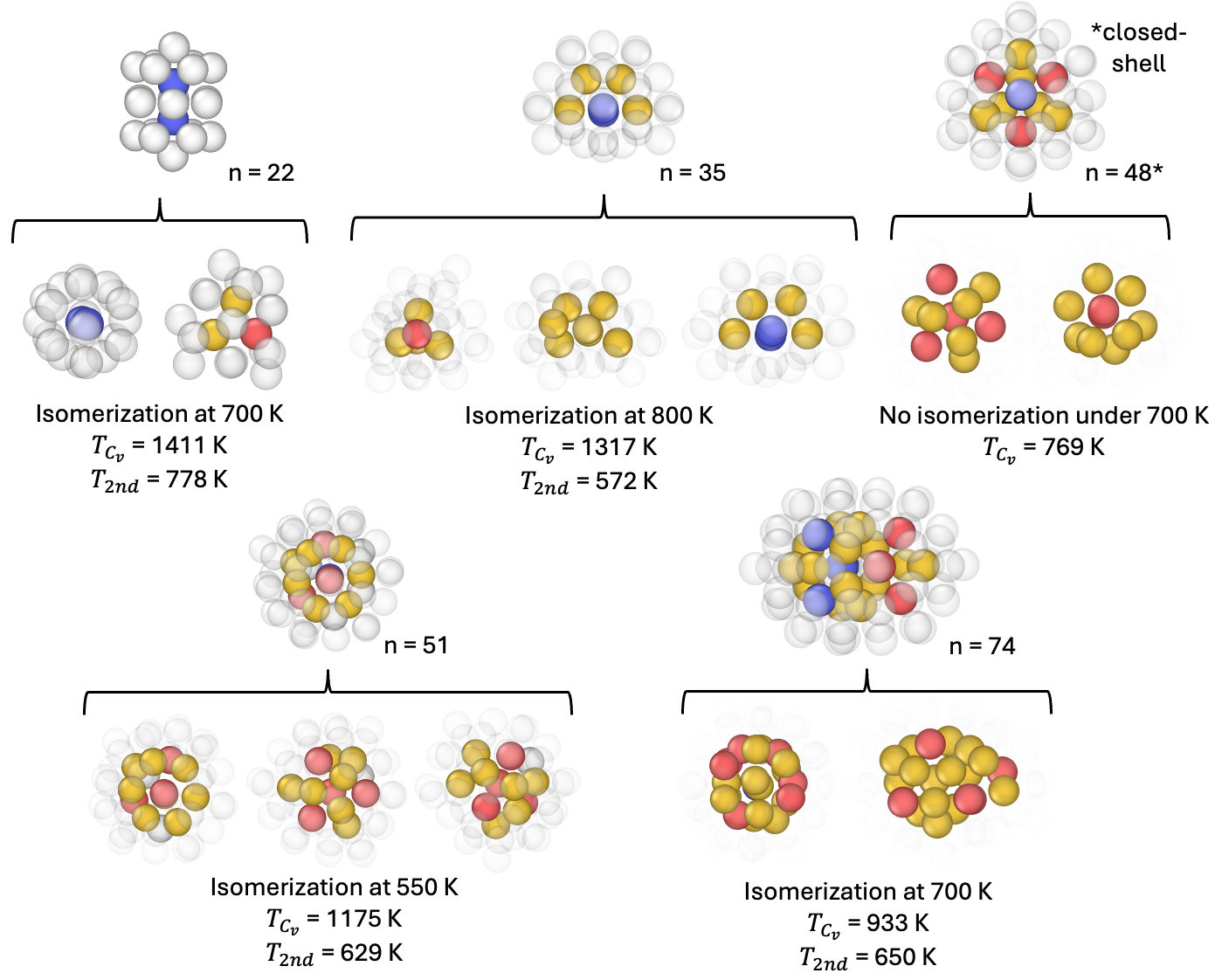}
\caption{Snapshots of global minimum energy structures (above) and frequently visited isomers for several cluster sizes. Second-order-like and near-closed-shell $Fe_{22}$ ($Fe_{23}$ is closed-shell), second-order-like and near-closed-shell $Fe_{35}$ ($Fe_{34}$ is closed-shell and second-order-like), closed-shell $Fe_{48}$, and second-order-like $Fe_{51}$ and $Fe_{74}$. The second-order-like clusters exhibit isomerization between structures several hundred kelvin below $T_{C_v}$.}
\label{fig: struct_competition}
\end{figure}

To understand the second-order-like melting process, trajectory files of NVT simulations at many temperatures and cluster sizes were examined to determine the lowest temperature at which isomerization between different structures occurred for a significant proportion of simulation time. Snapshots of global minimum energy configurations and structural isomers at temperatures where they appear for appreciable time are given in Fig.\ref{fig: struct_competition}.
Clusters $Fe_{22}$ and $Fe_{35}$ exhibited second-order-like melting curves (Cv-plots folder in SI), with high $T_{C_v}$s of 1411 K and 1317 K, respectively. The upper left 3 structures in Fig.\ref{fig: struct_competition} are snapshots of the global minimum configuration of $Fe_{22}$ and two competing structural isomers (a 2-atom linear core structure and a 3-atom planar core structure) that begin frequent interchange at 700 K—less than half of the temperature of maximum $C_v$. Simulations at higher temperatures show a lack of preferred core structure. The $Fe_{35}$ cluster fluctuates between 3 isomers at 800 K, far below its $T_{C_v}$ of 1317 K. The $Fe_{35}$ isomer cores (left to right) consist of a small tetrahedral core, a mostly-formed 5-atom ring structure with an extra atom to form a ring-tetrahedron amalgam, and the mostly-formed 5-atom ring structure alone, which is shared by the global minimum energy configuration.
The closed-shell first-order melting $Fe_{48}$ cluster was investigated to compare with smaller second-order-like melters. $Fe_{48}$ exists in the midsize region of cluster space with many structures of similar energies. As such, it exhibits isomerization between an anti-Mackay icosahedral core (akin to its global minimum configuration) and a 6-atom ring core with 2 extra core atoms (upper right 3 structures in Fig.\ref{fig: struct_competition}). The interchange between these isomers does not begin until 700 K, and becomes frequent at 750 K, just under its $T_{C_v}$ of 769 K.
Clusters with 51 and 74 atoms, which also have second-order-like melting behavior, also show frequent interchange between competing isomers (lower cluster snapshots in Fig.\ref{fig: struct_competition}). $Fe_{51}$ (left 4 structures) has frequent interchange at 550 K between partially filled Mackay (akin to its global minimum structure) and anti-Mackay icosahedral cores. This isomerization occurs well under the cluster's $T_{C_v}$ of 1175 K. $Fe_{74}$, in simulations at 700 K, shifts between filled trigonal-symmetric Mackay cores with island growth (like its global minimum configuration) and a compact version of the double-2-shell icosahedral structure (seen in $Fe_{78}$). Larger 2-shell clusters exhibiting second-order-like transitions don't have as high $T_{C_v}$ values as the smaller 2-shell clusters mentioned earlier. Still, the isomerization between these structures occurs over 200 K below its $T_{C_v}$ of 933 K.
All clusters analyzed with second-order-like melting behavior occur at sizes with 2 or more competing structures, and these clusters exhibit isomerization between competing isomers at temperatures well below $T_{C_v}$. Conversely, first-order-like cluster sizes do not show structural isomerization until temperatures near their $C_v$ peak. This phenomenon suggests that restructurings between competing structures at lower temperatures induce second-order-like melting behavior. With a mapped-out potential energy landscape of a particular cluster size, one could therefore predict that a size with multiple isomers of similar energy separated by a small barrier will have second-order-like melting transitions.

\subsection{Atomic mobility within clusters}
To investigate the mechanisms of cluster phase transitions, the cluster Lindemann index\cite{lindemann1910} ($\delta_c$) was calculated from each NVT MD trajectory. $\delta_c$ is often used to quantify the melting of clusters because it measures changes in atomic mobility in a solid, rapidly increasing upon initiation of atom movement. The temperature at which $\delta_c$ increases sharply is defined as the cluster's surface melting temperature, $T_{\delta_c}$. Melting points derived from $\delta_c$ are often defined as the temperature where the index meets an arbitrary threshold, such as 0.1 or 0.2\cite{frantz_magic_2001}. Because no threshold is widely used in nanocluster systems, this work picks the temperature just below which $\delta_c$ rises notably. The Lindemann index metric is sensitive to quasi-ergodicity in MD simulations\cite{frantz_magic_2001}, and some cluster sizes may be overestimated $T_{\delta_c}$ due to poor convergence (despite the long 20 ns simulation times).

\begin{figure}
    \centering
    \includegraphics[width=0.9\linewidth]{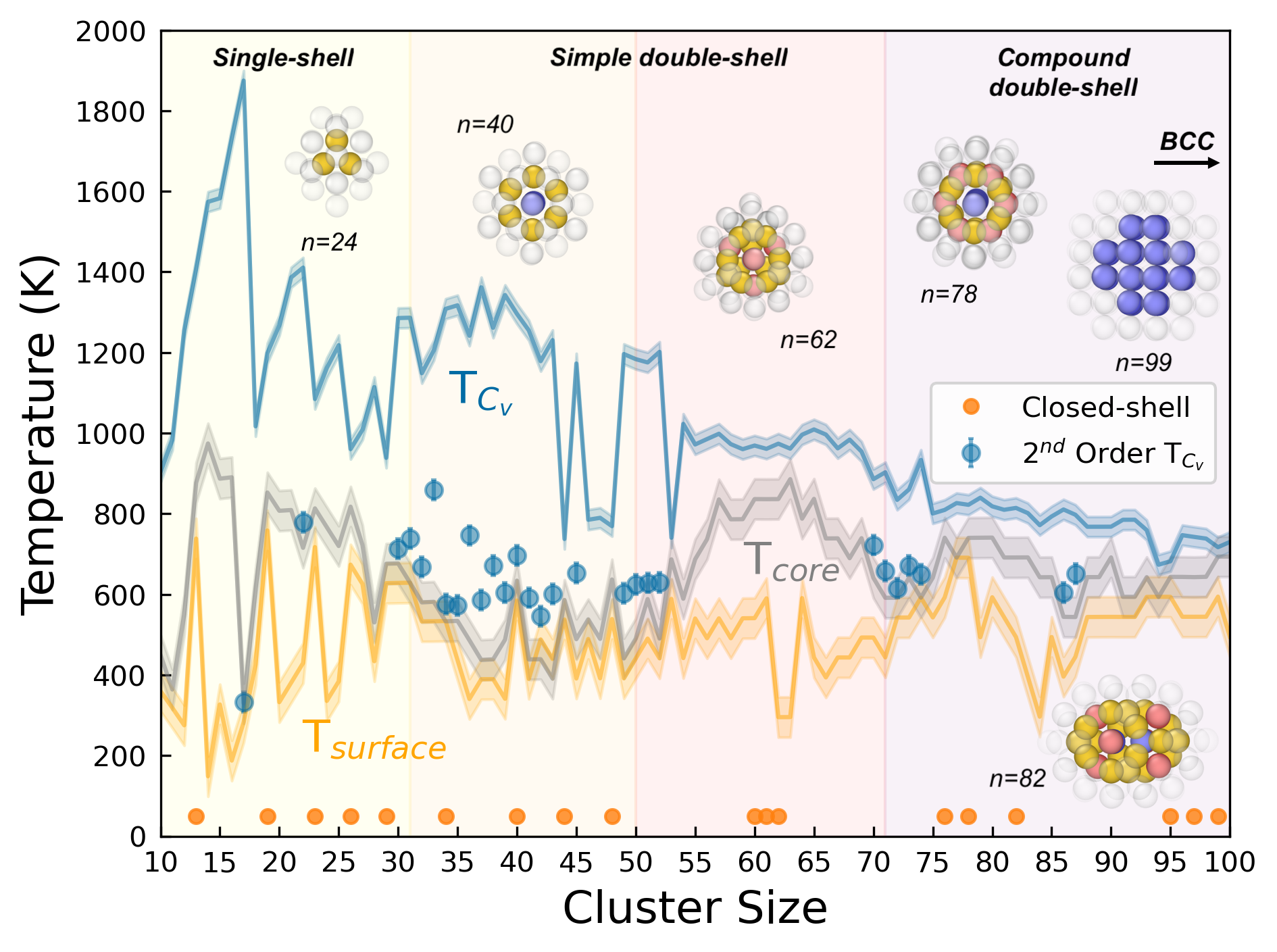}
\caption{Plot of melting temperatures $T_{C_v}$ (blue), $T_{core}$ (gray), and $T_{\delta_c}$ (orange), as a function of cluster size. Closed-shell clusters are indicated with orange points, and sizes with second-order-like phase transitions have their $T_{2nd}$ plotted in blue below the $T_{C_v}$ curve. Cluster structure regimes are denoted with colored fills in the body of the plot: 1-shell (yellow), Simple 2-shell (orange for ring-type cores and red for trigonal-symmetric Mackay cores), Compound 2-shell (purple), and bcc-dominant clusters are noted to lie at larger sizes. Cluster snapshots are placed in their given structure regimes.}
\label{fig: Tcv_Tdelta_Tcore_fig}
\end{figure}

For 1-shell cluster sizes ($n < 31$ atoms), closed-shell clusters have much higher $T_{\delta_c}$ than their immediate nearest neighbors (e.g., the surface of $Fe_{19}$ melts at 900 K, whereas $Fe_{20}$'s surface melts at 350 K – see Fig.\ref{fig: Tcv_Tdelta_Tcore_fig}). Larger near-closed-shell clusters do not have significantly lower $T_{\delta_c}$ than their corresponding closed-shell sizes.

\begin{figure}
\centering
  \includegraphics[width=0.7\linewidth]{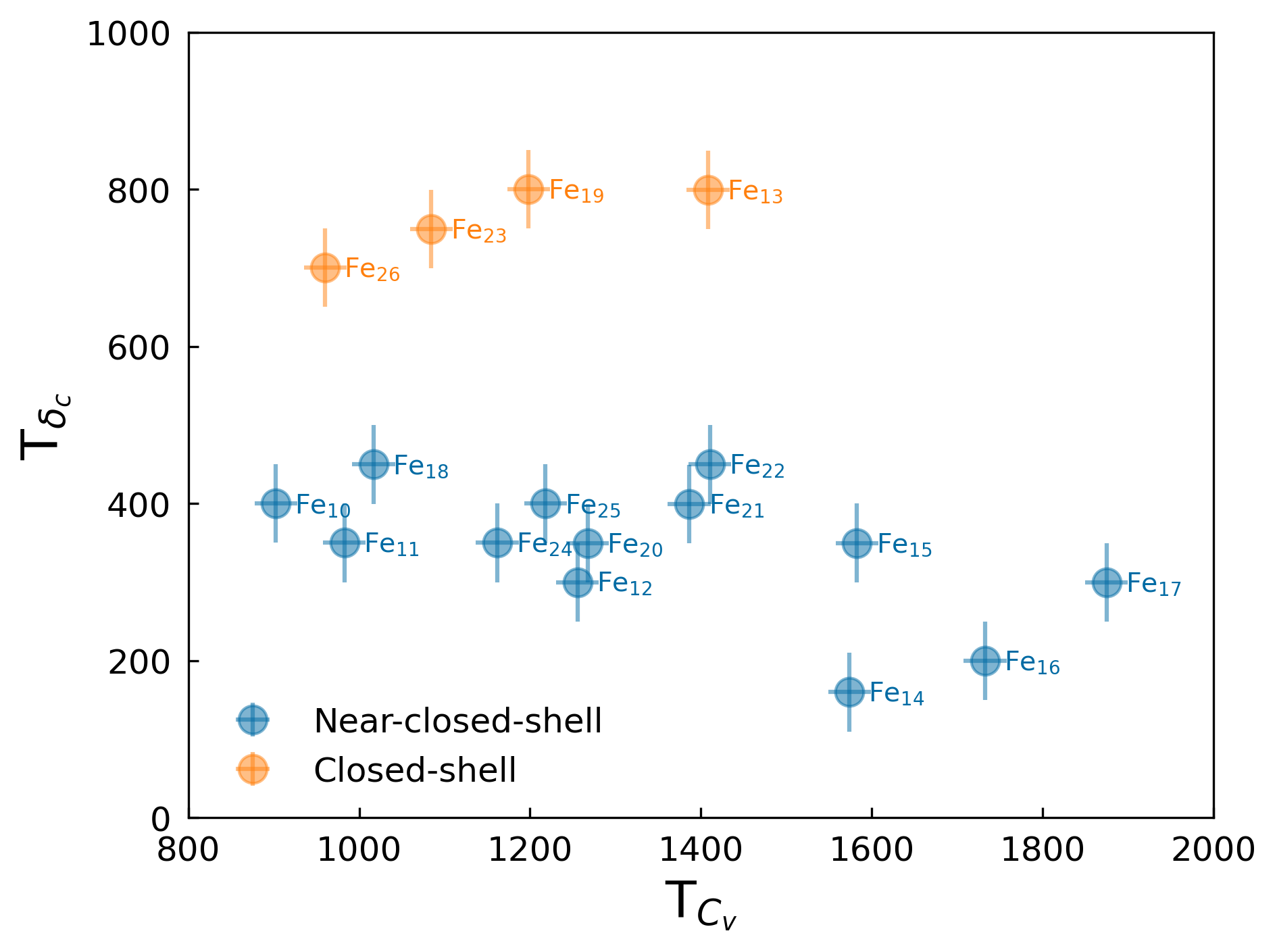}
\caption{Scatter plot of $T_{C_v}$ and $T_{\delta_c}$ for small cluster sizes. Near-closed shell clusters (blue) have lower surface melting temperatures $T_{\delta_c}$ than closed-shell clusters (orange).}
\label{fig: scatterplot_fig}
\end{figure}

\begin{figure}
\centering
  \includegraphics[width=0.7\linewidth]{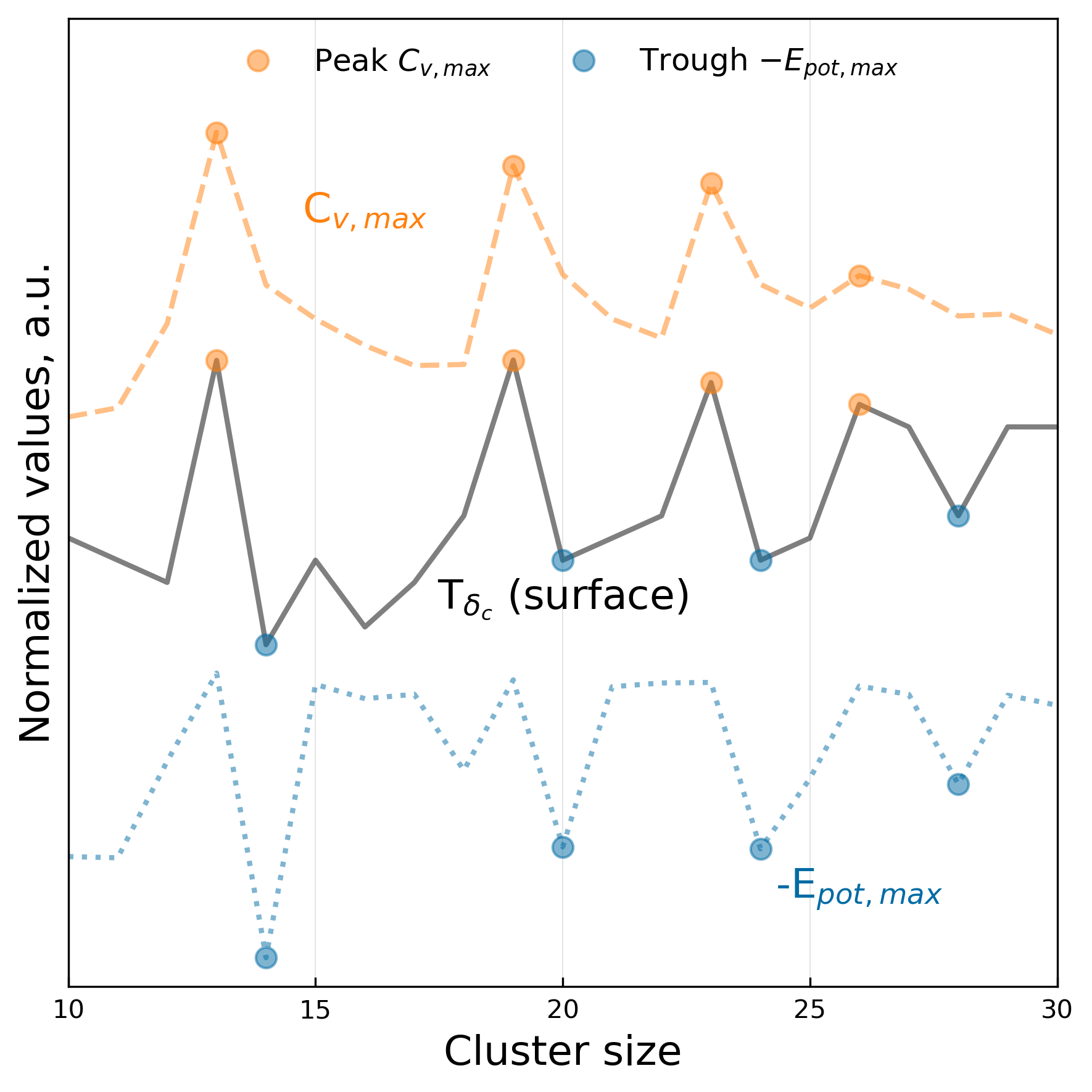}
\caption{$T_{\delta_c}$ (gray), $C_{v,max}$ (orange), and negative atomic $E_{pot,max}$ for minimum energy clusters with 30 or fewer atoms. Peaks in surface melting point coincide with peaks in $C_{v,max}$ (closed-shells). Troughs in surface melting point are predicted by the presence of a low negative $E_{pot,max}$ (a highly undercoordinated surface atom).}
\label{fig: maxCV_maxPE_fig}
\end{figure}

The phenomenon for small clusters can be observed on the scatterplot of $T_{C_v}$ and $T_{\delta_c}$ in Fig.\ref{fig: scatterplot_fig}. Surface melting of clusters 1, 2, or 3 atoms away from closed-shell structures is several 100s of K lower than closed-shell surface melting.
One can assume that clusters with one extra atom atop a closed shell have lower $T_{\delta_c}$ due to the undercoordinated atom having a lower barrier to surface diffusion. The highest potential energy on a single atom was recorded for each cluster size's global minimum energy configuration ($E_{pot,max}$) and compared to $T_{\delta_c}$ in Fig.\ref{fig: maxCV_maxPE_fig}.
For cluster sizes under 30 atoms (1-shell clusters), a peak in $C_{v,max}$ is a good indicator that a cluster size has a high $T_{\delta_c}$ and a trough in negative $E_{pot,max}$ (a peak in $E_{pot,max}$) indicates a low $T_{\delta_c}$. The absence of a trough in $T_{\delta_c}$ for $Fe_{18}$ may be explained by the fact that $Fe_{19}$ has a closed-shell structure, making the $Fe_{18}$ structure—which is the same structure as $Fe_{19}$ minus one atom—more robust despite its high maximum potential energy atom.

\begin{figure}
\centering
\begin{minipage}{.25\linewidth}
  \centering
  \includegraphics[width=\linewidth]{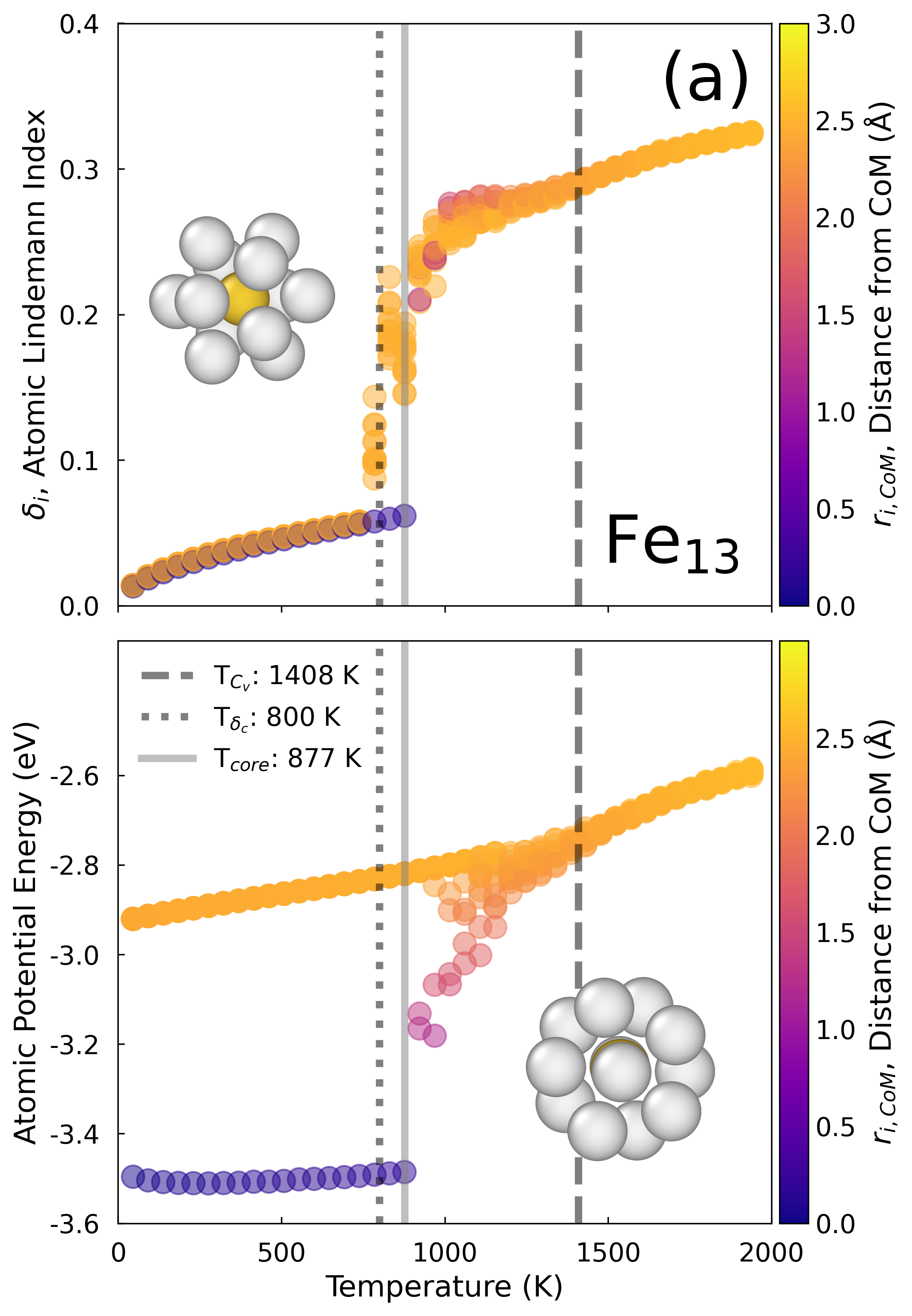}
\end{minipage}
\begin{minipage}{.25\linewidth}
  \centering
  \includegraphics[width=\linewidth]{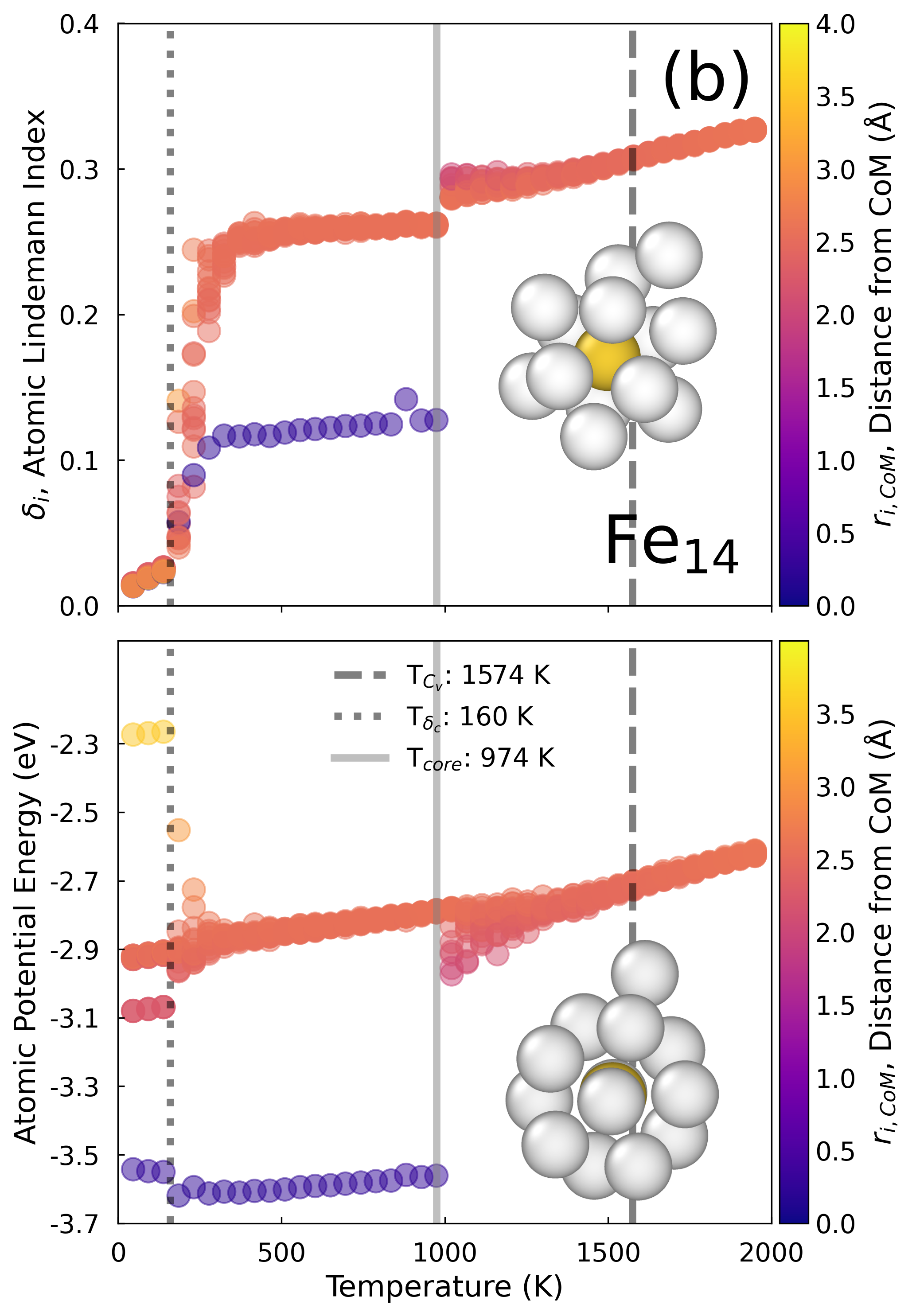}
\end{minipage}
\begin{minipage}{.25\linewidth}
  \centering
  \includegraphics[width=\linewidth]{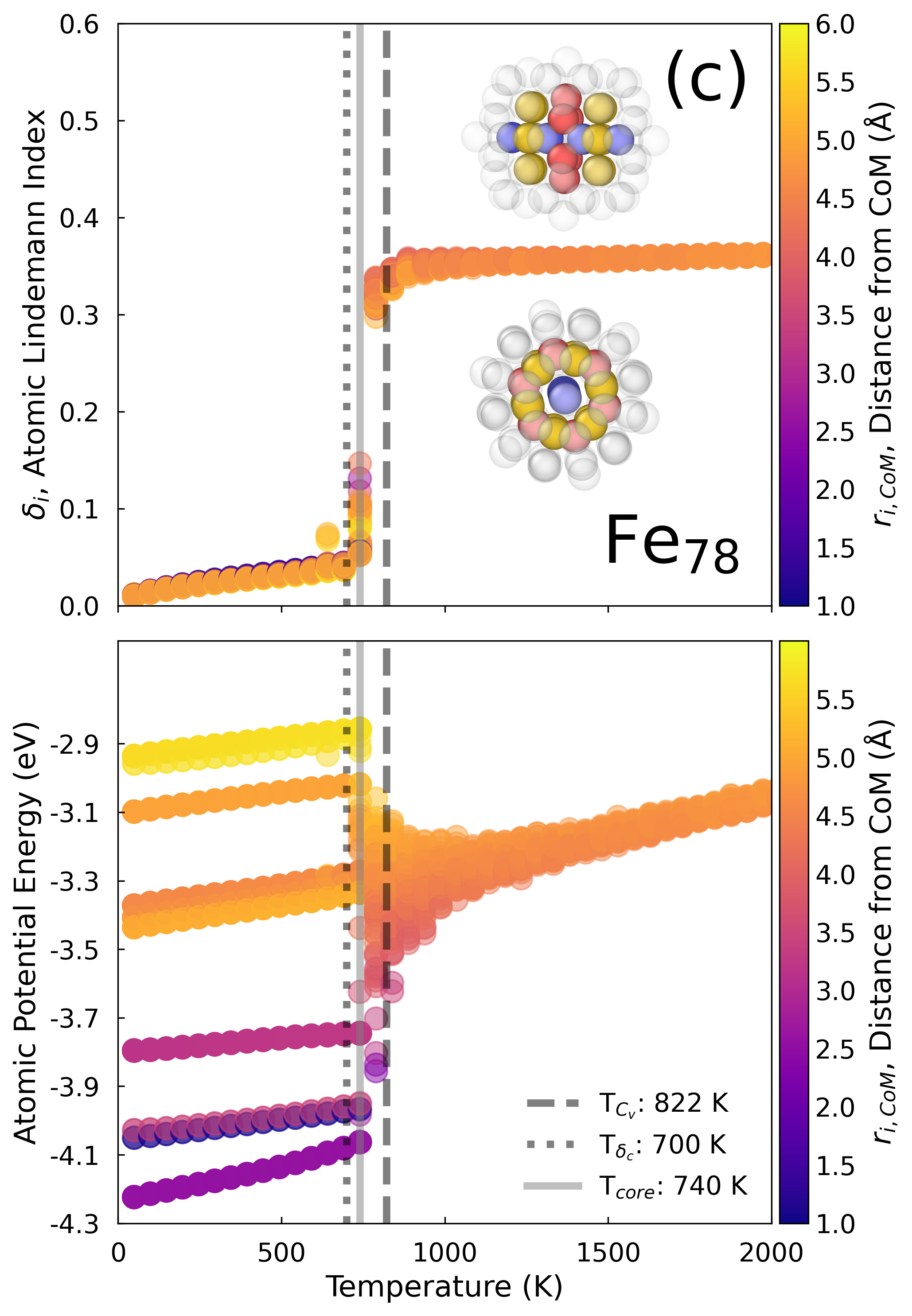}
\end{minipage}
\begin{minipage}{.25\linewidth}
  \centering
  \includegraphics[width=\linewidth]{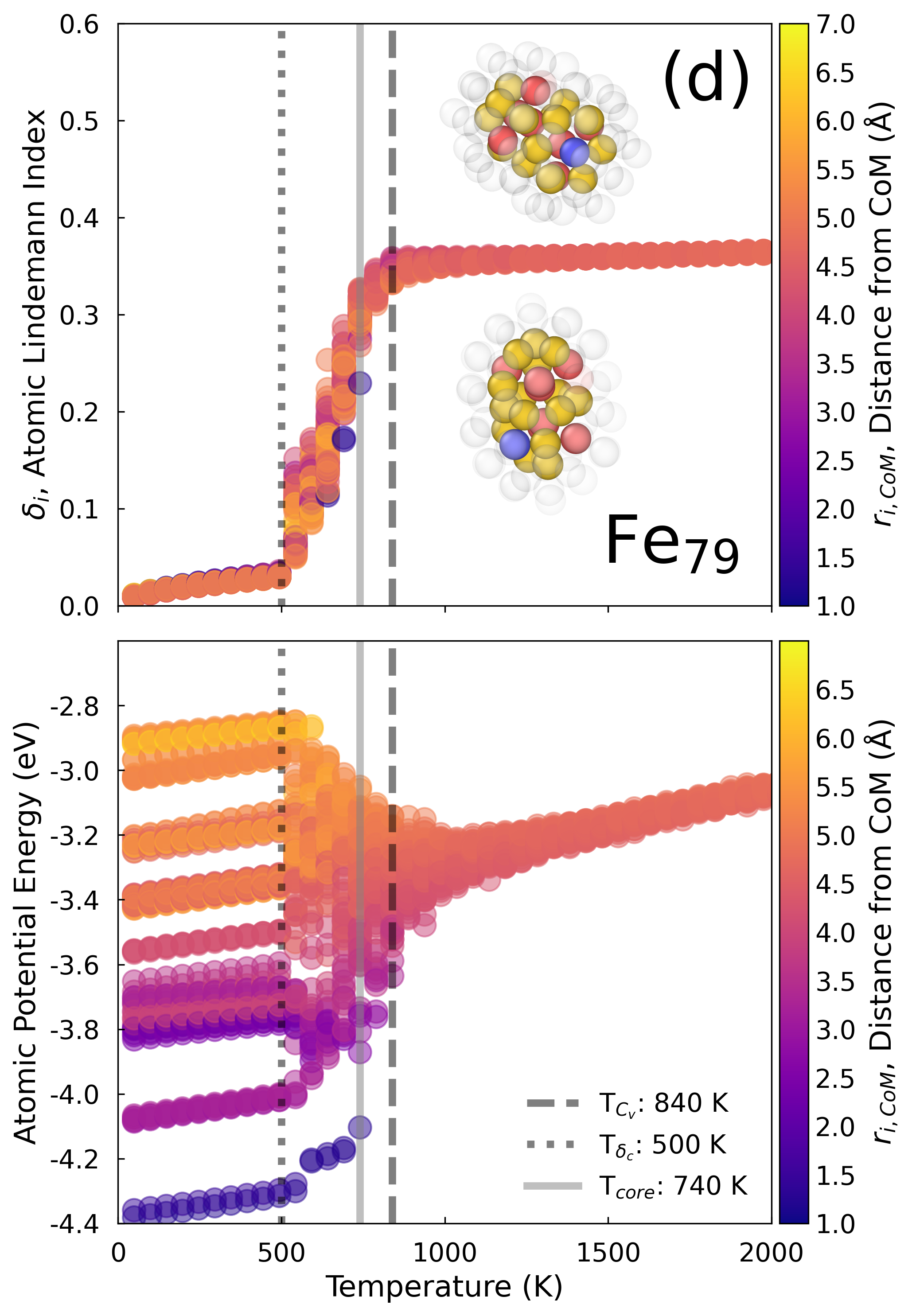}
\end{minipage}
\caption{Plots of atomic Lindemann indices, $\delta_i$ and atomic potential energy, colored by atomic distance from cluster CoM, $r_{i,CoM}$, as a function of temperature (colored by $\delta_c$ or $r_{i,CoM}$) for \textbf{(a)} closed-shell $Fe_{13}$ and \textbf{(b)} near-closed $Fe_{14}$, and \textbf{(c)} closed shell $Fe_{78}$ and \textbf{(d)} near-closed $Fe_{79}$. Surface melting, $T_{\delta_c}$, $T_{C_v}$, and core melting, $T_{core}$, are indicated on each plot.}
\label{fig: individual_delta_r_fig}
\end{figure}

For more detailed mechanistic information on atom position and mobility during melting, atomic Lindeman indices, $\delta_i$, average atomic distance from cluster center-of-mass (CoM), $r_{i,CoM}$, and average potential energy of all atoms, were calculated from the NVT simulation trajectories at each temperature. Selected 2D projections of $\delta_i$, $r_{i,CoM}$, atomic potential energy, and temperature are shown for several cluster sizes in Figs.\ref{fig: individual_delta_r_fig}a-d.
In the top plots for Figs.\ref{fig: individual_delta_r_fig}a-d, $\delta_i$ is plotted as a function of temperature and colored by atomic distance from the cluster CoM (purple is closer to the center, yellow is closer to the surface). In the bottom plots, atomic potential energy is a function of temperature, colored again by $r_{i,CoM}$. Note that transparent points stacked on top of each other can appear darker despite the color of the markers being the same.

The use of $r_{i,CoM}$ and atomic potential energy allows for the visualization of the cluster shells and measurement of core melting (interlayer mixing). Using this value, the core melting temperature, $T_{core}$, was defined as the temperature at which the closest atom to the center of mass begins to become mobile. For both $Fe_{13}$ and $Fe_{14}$ (Figs.\ref{fig: individual_delta_r_fig}a-b), the cluster cores (the inner atom) become mobile at much higher temperatures than the surface atoms. Because $Fe_{14}$ is small and near-closed, its extra surface atom becomes mobile at a low temperature of 160 K, but its core remains unmoved until 974 K, a higher temperature than it takes to melt both the surface and core of $Fe_{13}$! While not present in $Fe_{78}$ (Fig.\ref{fig: individual_delta_r_fig}c), surface melting before the core was commonly observed for larger clusters, with differences between $T_{\delta_c}$ and $T_{core}$ of 100 K to several 100s of K. This is the case with $Fe_{79}$ (Fig.\ref{fig: individual_delta_r_fig}d). Despite the surface melting before the core in $Fe_{79}$, its core melts around the same temperature as $Fe_{78}$ at 740 K. The mechanistic 2D projection plots for all cluster sizes examined are available in the delta-r-pe-plots folder in SI.

All melting points ($T_{C_v}$, $T_{\delta_c}$, and $T_{core}$) are plotted as a function of cluster size in Fig.\ref{fig: Tcv_Tdelta_Tcore_fig} to analyze the variation in melting properties over the whole size range studied.
In general, $T_{\delta_c} < T_{core} < T_{C_v}$. This is expected, as surface melting occurs at lower or similar temperatures as core melting, and maximum fluctuations in potential energy are expected to occur upon dissociation of a core structure (at higher temperatures than the initiation of core atom movement, $T_{core}$). Notably, most second-order-like cluster sizes show much higher $T_{C_v}$ than neighboring atoms (e.g., $T_{C_v}$ for $Fe_{49-53}$ is over 400 K above that of $Fe_{48}$ and $Fe_{54}$ despite having similar core structures). The second-order-like melting clusters occur frequently in the building of the second shell, and during sizes where large structural transitions take place in the minimum energy configurations ($Fe_{49-53}$ and $Fe_{70-74}$). $Fe_{86-87}$ also exhibit second-order-like melting at the transition between two connected trigonal-symmetric Mackay cores ($Fe_{82}$) and double-2-shell-icosahedra with island growth (similar to $Fe_{97}$ in Fig.\ref{fig: min_structs}).

For 1-shell cluster sizes ($n < 31$, yellow fill), $T_{C_v}$ and $T_{\delta_c}$ have opposing peaks and troughs at closed-shell cluster sizes: i.e.,  $T_{C_v}$ is smallest when $T_{\delta_c}$ is largest. Cluster sizes right after closed-shells have larger $T_{C_v}$ (as these clusters retain the stability of the closed-shell structure below) and smaller $T_{\delta_c}$ due to undercoordinated surface atoms having low barriers to diffusion (Fig.\ref{fig: maxCV_maxPE_fig}). $T_{core}$ is relatively smooth in this region, aside from a strong dip at $Fe_{17}$, which has a low-temperature isomerization event between a trigonal-symmetric structure with an overcoordinated central atom, and an icosahedral structure with island growth.

Ring-dominated simple 2-shell cluster sizes ($31 \leq n \leq 50$, orange fill) resist interpretation aside from high $T_{C_v}$s due to second-order-like melting (due to many competing isomers in the size range). $Fe_{44}$ and $Fe_{46-48}$ are exceptions in this range with stronger first-order-like transitions that explain their $T_{C_v}$. As mentioned earlier, $Fe_{34}$ and $Fe_{40}$ have weak first-order-like peaks at lower temperatures, so their $T_{C_v}$s high like the surrounding second-order-like cluster sizes, while these peaks are taken as their $T_{2nd}$s.

Trigonal-symmetric Mackay core structures dominate in larger simple 2-shell clusters ($50 \leq n \leq 71$, red fill). The range consists of second-order-like transitions early and late in the size range, where anti-Mackay core structures ($Fe_{49}$) and double-2-shell icosahedra ($Fe_{78}$) compete with the conserved trigonal-symmetric core. Near the middle of the size range, $T_{C_v}$ and $T_{core}$ are elevated due to the integrity of the trigonal-symmetric Makcay core. $Fe_{62-63}$ have depressed $T_{\delta_c}$ due to undercoordinated surface atoms.

Compound 2-shell cluster sizes ($72 \leq n \leq 100$, purple fill) have two sub-regimes: one regime has double-2-shell icosahedra ($Fe_{78}$) or joined-trigonal-symmetric Mackay cores ($Fe_{82}$), and the other regime has large island growth atop the double-2-shell icosahedral structure, culminating with $Fe_{97}$ (see Fig.\ref{fig: min_structs}). $Fe_{84}$ has a low $T_{\delta_c}$ due to a highly undercoordinated surface atom.

Three larger cluster sizes, $Fe_{95}$ and $Fe_{98-99}$, have global minimum structures with bcc structure. Longer 40 ns PT simulations were attempted for all clusters above 90 atoms, but only these 3 cluster sizes crystallized to the bcc structure.

\section{Discussion}
\label{discussion}
The analyses in this work illustrate the complexity of nanocluster melting, even in the narrow size range 
of under 100 atoms ($< 1.2$ nm). All cluster properties studied were sensitive functions of cluster size: the shape of the $C_v$ curve, surface melting point ($T_{\delta_c}$), core melting point ($T_{core}$), and energetic melting point ($T_{C_v}$ – the temperature of maximum $C_v$). Melting of clusters with one shell ($n \leq 30$ atoms) differs qualitatively from melting of 2-shell clusters with different types of core structures ($31 \leq n \leq 50$, and $51 \leq n \leq 71$), which differ still from melting in clusters with compound 2-shell structures ($72 \leq n \leq 100+$). Given the appearance of stable bcc chunks of Fe in the 90-atom range, the melting of these nanoclusters and larger likely compose a regime of their own, blending into nanoparticle behavior and Gibbs-Thomson melting point depression.

\subsection{Second-order-like phase transitions}
Clusters with sizes with two different core structures competing for stability showed reduced max($C_v$) in their heat capacity curves, elevated temperatures of maximum $C_v$ ($T_{C_v}$), and qualitatively distinct heat capacity curves akin to second-order phase transitions. Trajectories of the melting of these clusters showed isomerization between competing structures 100s of K below the $T_{C_v}$. In \textit{Frantz (2001)}\cite{frantz_magic_2001}, the phenomenon of depressed max($C_v$) was reported in medium-sized Lennard-Jones clusters for $Ar_{31-38}$, where the primary peak in $C_v$ is broadened and diminished, while a much sharper, lower-temperature peak arises, eventually rising above the original broad peak (the broad peak appears as a plateau into the liquid value of $C_v$). The $C_v$ curves in that work appear to be second-order-like until the secondary low-temperature peak rises above the $C_v$ of the liquid. The authors attribute the appearance of the small peaks at $Ar_{31}$ to a restructuring in the minimum energy configurations from anti-Mackay to Mackay icosahedral overlayers. They note a high density of structural isomers of similar energy in this size range\cite{Doye_evolution_of_potential}, observed in other studies using a Morse potential\cite{Doye_effect_of_the_range}. It is unclear why the current work does not observe the sharp low-temperature peaks in $C_v$ plots for sizes aside from $Fe_{34}$ and $Fe_{40}$.

Second-order phase transitions in clusters have been reported experimentally and in simulation in cases where the system undergoes a semi-conducting to metallic transition or shift in magnetic moment\cite{breaux_second-order_2005,Experimental_specific_Gerion}, neither of which are modeled in the classical MD potential in this work. For this reason, the second-order-like nature of the cluster phase transitions in this work must come only from geometric structural factors.

\subsection{1-shell clusters: $n < 31$}
For nanoclusters under 30 atoms, melting begins with surface atoms swapping places with each other ($T_{\delta_c} < T < T_{core}$). Eventually, the cluster's central atom(s) breaks free ($T_{core} < T < T_{C_v}$), and the cluster forms different structures with higher potential energies (liquid) and short lifetimes. At higher temperatures, the higher potential energy structures become longer-lived ($T > T_{C_v}$), eventually pushing the cluster into the disordered liquid state.

As expected from previous studies\cite{bagrets_lowering_2010}, near-closed shell and closed-shell clusters maintain similar core melting temperatures ($T_{core}$), but near-closed shell clusters have much lower surface melting points ($T_{\delta_c}$) than neighboring closed-shells due to the presence of under-coordinated atoms on the cluster surface (Fig.\ref{fig: maxCV_maxPE_fig}). It was shown that peaks in the potential energy of the least stable atom in the global minimum cluster configurations (which coincided with clusters with 1 more atom than a closed-shell size) conferred sharply reduced $T_{\delta_c}$. Conversely, a peak in max($C_v$) conferred sharp increases in $T_{\delta_c}$.

Interestingly, the $T_{C_v}$ is elevated for clusters with 1 to a few more atoms than a closed-shell cluster. This was attributed to the retained stability from the closed shell core structure, which resisted melting at higher temperatures. The sizes after closed-shell clusters could store excess energy in the degrees of freedom of the extra surface atoms.

\subsection{Small 2-shell clusters: $31 \leq n \leq 50$}
Clusters above 30 atoms enter a regime where a 2-shell structure is possible with a 5-atom ring and 1 central atom (similar to $Fe_{34}$ in Fig.\ref{fig: min_structs}); however, this type of 2-shell structure competes with 1-shell structures with stable icosahedral cores. This competition is proposed to cause most clusters in this size range to melt in second-order-like fashion. Closed-shell $Fe_{44}$ and $Fe_{48}$ are exceptions to this trend, and have significantly lower $T_{C_v}$ than neighboring sizes as a result. The top part of the size range enters a regime where larger 6-atom ring inner shell structures compete with 1-shell anti-Mackay and Mackay icosahedral cores.

\subsection{Large 2-shell clusters: $51 \leq n \leq 71$}
While the lower end of this size range consists of many second-order-like melters, a consistently robust trigonal-symmetric Mackay icosahedral core structure gives many clusters a higher core melting temperature. This attribute yields similarity to nanoparticle melting in terms of surface melting before core melting (i.e., $T_{\delta_c} < T_{core}$). Clusters have a stable closed-shell core in common, with island structures accumulating on the surface with increasing size. The end of the size range begins to compete with 2-shell structures with multiple domains.

\subsection{Compound 2-shell and bulk-like clusters: $72 \leq n \leq 100$}
Due to the large number of atoms required to add a new stable core domain, competing structures only overlap with 2 cluster sizes in this size range: $Fe_{86}$ and $Fe_{87}$, both of which have second-order-like melting as a result. Double-2-shell structures like $Fe_{78}$ and $Fe_{82}$ in Fig.\ref{fig: min_structs} compete with compound 2-shell clusters with $Fe_{78}$ cores and large island domains on top ($n > 87$). Due to the stable core structures and surface melting before core melting, clusters above $Fe_{87}$ exhibit less variation in $T_{C_v}$, $T_{\delta_c}$, and $T_{core}$, and all exhibit first-order-like melting. Bcc-structured clusters are energetically favorable for some cluster sizes; however, the barriers to isomerization between compound 2-shell structures and bcc are high, which was an issue for minimum energy configuration determination (even with enhanced sampling from parallel tempering). Fluctuation between these structures either does not occur below the temperature of cluster melting or occurs on timescales larger than those practical with classical MD. The presence of these bcc clusters indicates a transition into larger nanoparticle melting point scaling (Gibbs-Thomson) outlined in the Introduction.

\subsection{Potential implications for catalysis}
The addition of one to two atoms in a cluster can cause strong variations in melting points of the surface ($T_{\delta_c}$) and core ($T_{core}$), as well as the `energetic' melting point ($T_{C_v}$, temperature with maximum $C_v$). Size-dependent melting behavior may confer size-dependent activity in catalytic nanocluster systems at temperatures where some clusters are melted and others are not. Differences in surface and core melting can influence carbon adsorption, diffusion, and dissolution in CNT growth, leading to different growth rates or modes of CNT growth (tangential vs. perpendicular growth\cite{amara_modeling_2017}). Surface atom mobility ($T > T_{\delta_c}$) could accelerate the diffusion of C atoms on the NP surface, increasing the CNT growth rate. Core atom mobility ($T > T_{core}$), however, could increase the solubility of carbon in Fe, requiring more C to be fed to the NPs and thus slowing the CNT growth rate. An increase in potential energy (low $T_{C_v}$) associated with longer lifetimes of disordered isomers – with a disordered outer layer – may change both the adsorption energies and transport properties.

\section{Conclusion}
\label{conclusion}
The melting of Fe nanoclusters up to 100 atoms ($< 1.2$ nm) in size was investigated with classical molecular dynamics simulations.

The key takeaways from this work are the following:
\begin{itemize}
    \item The addition of one to two atoms in a cluster can cause strong variations in melting points of the surface ($T_{\delta_c}$) and core ($T_{core}$), as well as the `energetic' melting point ($T_{C_v}$, temperature with maximum $C_v$). There may be implications for enhanced or suppressed catalytic activity through changes in species adsorption or transport on the cluster surface.
    \item Second-order-like melting behavior, defined by the absence of a distinctive peak in the heat capacity curve, is observed for many cluster sizes where the most stable core structures are very close in energy. Due to the small number of atoms in atomic clusters, the barriers to transition between these structures are low, allowing for isomerization at low temperatures. These cluster sizes have high $T_{C_v}$s, low $T_{\delta_c}$s, and low $T_{core}$s.
    \item Second-order phase transitions in clusters have only previously been reported in systems with concurrent transitions in electronic or magnetic behavior\cite{wallace_melting_1997,breaux_second-order_2005} not modeled in the classical interatomic potential used in this work.
    \item  1-shell clusters ($n < 31$) with closed shells have high surface melting temperatures $T_{\delta_c}$, while clusters with 1 to a few more atoms than their neighboring closed-shell cluster have reduced $T_{\delta_c}$ and increased $T_{C_v}$ due to the cluster's closed-shell internal core.
    \item Small 2-shell clusters ($31 \leq n \leq 50$) have notoriously high densities of structures with similar energies, leading to many second-order-like melting clusters.
    \item Many larger 2-shell clusters ($51 \leq n \leq 71$) have a stable core structure, causing higher $T_{core}$, and surface melting before core melting.
    \item Compound 2-shell clusters ($72 \leq n \leq 100$) have multiple stable core domains, surface melting before core melting, and relatively few cluster sizes with second-order-like melting. Some cluster sizes ($n = 95,98,99$) prefer bcc core structures, trending towards Gibbs-Thomson nanoparticle melting point scaling.
\end{itemize}

\section*{Supplementary Materials}
\label{supplementary_materials}
Additional materials such as data used in method benchmarking can be found in \textbf{SI.pdf}. Heat capacity curves, parallel tempering acceptance ratio plots, Lindemann index curves, atomic Lindemann index vs. distance from CoM plots for all cluster sizes, and snapshots of minimum energy configurations for many cluster sizes can be found in the \textbf{Cv-plots}, \textbf{Pacc-plots}, \textbf{lind-plots}, \textbf{delta-r-pe-plots}, \textbf{min-structs} folders, respectively.

\section*{Acknowledgments}
\label{acknowledgements}
The support of Princeton University's Andlinger Center for Energy and the Environment, and the Program in Plasma Science and Technology at the Princeton Plasma Physics Laboratory is gratefully acknowledged. In addition, this research utilized computing resources on the Princeton University Della and Stellar clusters. We also thank the anonymous reviewers for detailed questions and suggestions that helped to improve the paper.\\

\textbf{COI statement}: The authors have no conflicts of interest to disclose.\\

\textbf{Data availability statement}: The data are contained within the article and supplementary material.\\

\textbf{Author contribution statement}: \\
\textbf{Louis E. S. Hoffenberg}: conceptualization (equal); formal analysis (lead); writing – original draft preparation (lead). \textbf{Alexander Khrabry}: conceptualization (equal); formal analysis (supporting); review and editing (equal). \textbf{Yuri Barsukov}: conceptualization (equal); review and editing (equal). \textbf{Igor D. Kaganovich}: funding acquisition (supporting); conceptualization (equal); supervision (equal). \textbf{David B. Graves}: funding acquisition (lead); conceptualization (equal); review and editing (equal); supervision (equal).

\bibliography{main}

\end{document}


\title{Supplementary Information for Types of Size-Dependent Melting in Fe Nanoclusters: a Molecular Dynamics Study}

\author{Louis E. S. Hoffenberg}
    \affiliation{Department of Chemical and Biological Engineering, Princeton University, Princeton, New Jersey 08540}
\author{Alexander Khrabry}
    \altaffiliation{Andlinger Center for Energy and the Environment, Princeton University, Princeton, New Jersey 08540}
\author{Yuri Barsukov}
    \altaffiliation{Princeton Plasma Physics Laboratory, Princeton, New Jersey 08540}
\author{Igor D. Kaganovich}
    \email{ikaganov@pppl.gov}
    \altaffiliation{Princeton Plasma Physics Laboratory, Princeton, New Jersey 08540}
\author{David B. Graves}
    \email{dgraves@princeton.edu}
\affiliation{Department of Chemical and Biological Engineering, Princeton University, Princeton, New Jersey 08540}

\date{\today}

\maketitle
\begin{figure}[h!]
    \centering
    \includegraphics[width=0.45\textwidth]{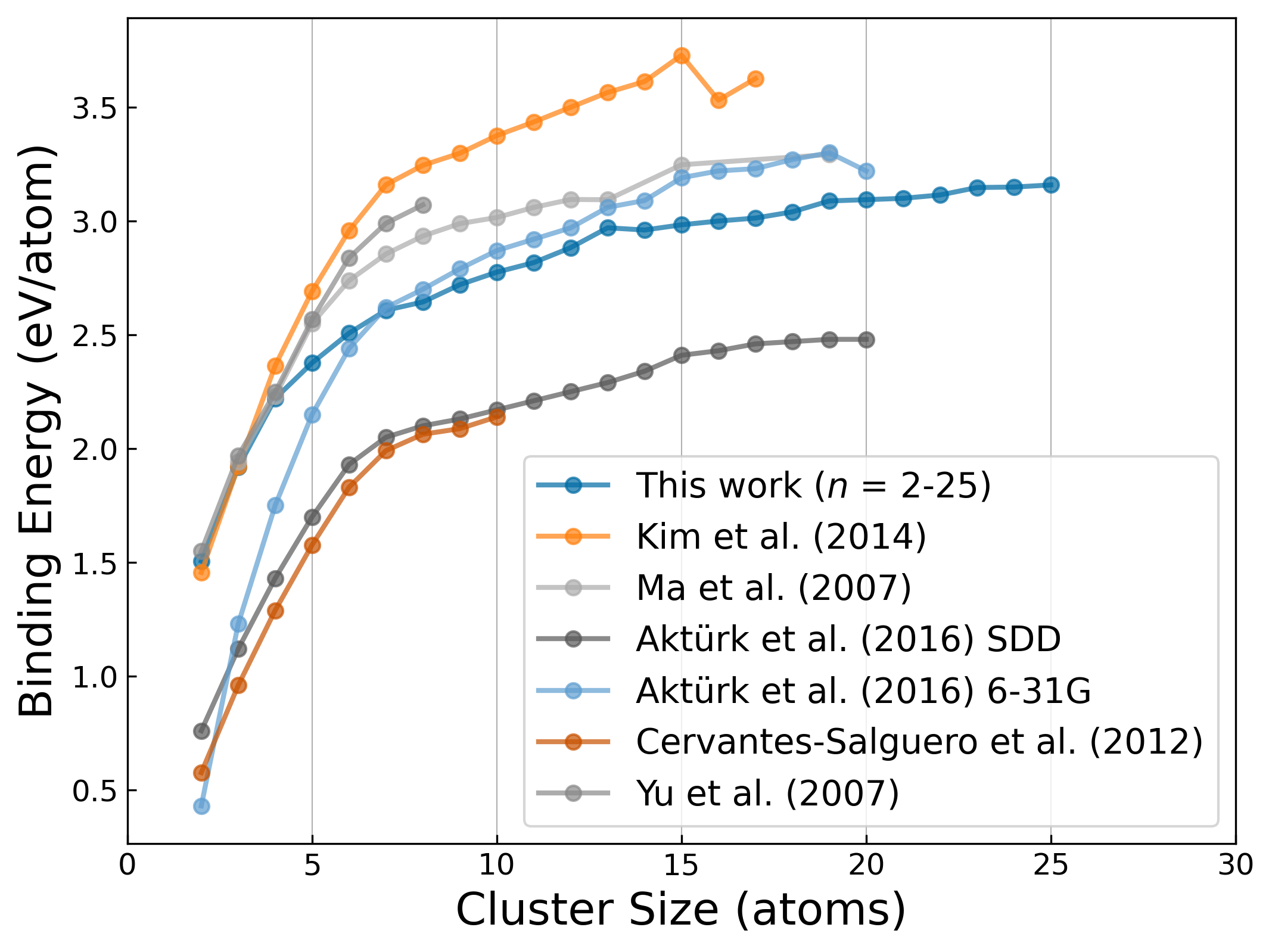}
    \caption{Plot comparing binding energies (BEs) from the EAM-FS classical MD potential with BEs from DFT calculations from several works\cite{kim_magic_2014, ma_structures_2007,akturk_bh-dftbdft_2016,cervantes-salguero_structure_2012,yu_theoretical_2007}. The EAM-FS potential yields BEs among those from the DFT literature.}
    \label{fig: DFT_BEs_comparison}
\end{figure}

\begin{figure}[h!]
    \centering
    \includegraphics[width=0.45\textwidth]{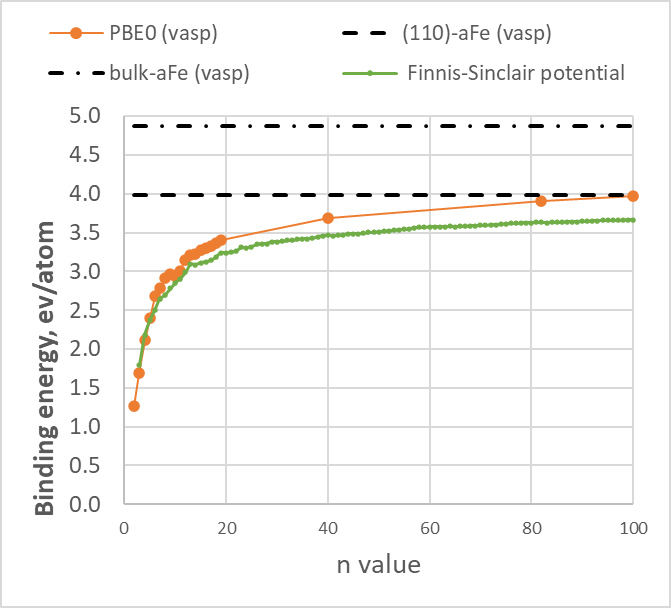}
    \caption{Plot comparing binding energies (BEs) from the EAM-FS classical MD potential with BEs from PBE0 DFT calculations for Fe clusters. Bulk $\alpha$-Fe and (110) $\alpha$-Fe binding energies are given as dash-dot and dashed lines, respectively. The EAM-FS potential (green) yields BEs close to that of PBE0 (orange) for many cluster sizes.}
    \label{fig: PBE0_BEs_DFT_comparison}
\end{figure}

\begin{figure}[h!]
    \centering
    \includegraphics[width=0.45\textwidth]{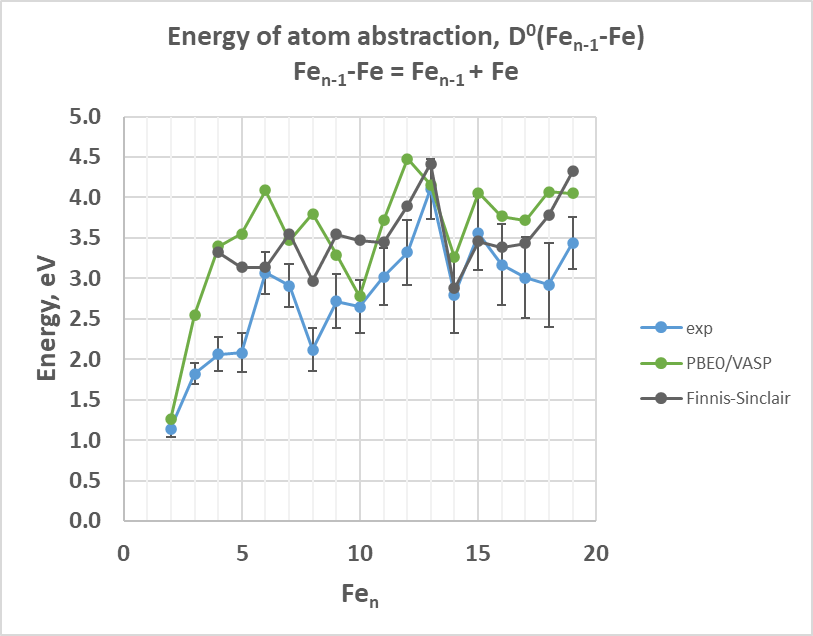}
    \caption{Atomic abstraction energies for Fe clusters up to 19 atoms with the EAM-FS classical MD potential, experiments\cite{lian_collisioninduced_1992}, and PBE0 DFT calculations. The EAM-FS potential (gray) yields abstraction near or between that of PBE0 (green) and experiments (blue) for many cluster sizes.}
    \label{fig: PBE0_E_abstraction_DFT_comparison}
\end{figure}

\begin{figure}[h!]
    \centering
    \includegraphics[width=0.45\textwidth]{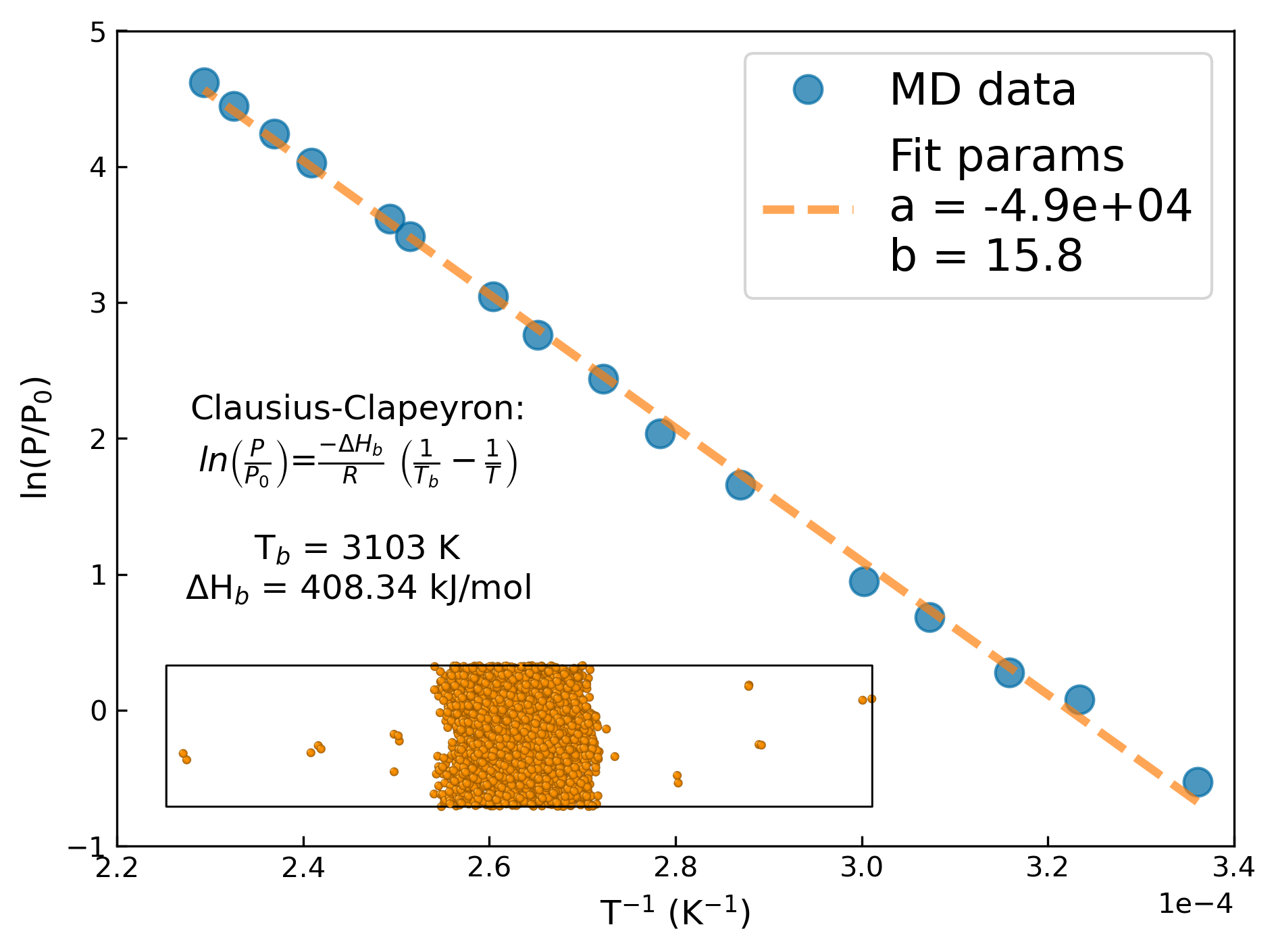}
    \caption{Plot of vapor pressure and temperature data from MD vapor-liquid NVE coexistence simulations. Fitting with the Clausius-Clapeyron relation allows for calculation of boiling point ($T_b$) and latent heat of boiling ($\Delta H_b$), close to experimental values for Fe: $3135$ K and $3.5\times10^{2}$ kJ/mol. A snapshot of an MD coexistence simulation at 4000 K is shown in the bottom left.}
    \label{fig: coexistence_VLE_data}
\end{figure}

\clearpage

\bibliography{SI}